\documentstyle[epsfig]{article}
\def\gappeq{\mathrel{\rlap {\raise.5ex\hbox{$>$}}
{\lower.5ex\hbox{$\sim$}}}}

\def\lappeq{\mathrel{\rlap{\raise.5ex\hbox{$<$}}
{\lower.5ex\hbox{$\sim$}}}}

\def\Toprel#1\over#2{\mathrel{\mathop{#2}\limits^{#1}}}

\begin{document}
\pagestyle{empty}
\begin{flushright}
hep-ph/0106221\\
Revised version
\end{flushright}
\vspace*{3mm}
\begin{center}
{\Large \bf
Fixation of   theoretical  ambiguities\\ in the improved fits
to the $xF_3$ CCFR data \\
at the next-to-next-to-leading order and
beyond}\\
\vspace{0.1cm}
{\bf A.L. Kataev$^{a}$, G. Parente$^{b}$ and A.V. Sidorov$^{c}$}\\
\vspace{0.1cm}
$^{a}$
Institute for Nuclear Research of the Academy of Sciencies of
Russia,\\ 117312, Moscow, Russia\\
\vspace{0.1cm}
$^{b}$ Department of Particle Physics, University
of Santiago de Compostela,\\ 15706 Santiago de Compostela, Spain\\
\vspace{0.1cm}
$^{c}$Bogoliubov Laboratory of Theoretical Physics, Joint
Institute for Nuclear Research,\\ 141980 Dubna, Russia\\
\end{center}
\begin{center}
{\bf ABSTRACT}
\end{center}
\noindent
Using the results for
the NNLO QCD corrections to anomalous dimensions
of odd $xF_3$ Mellin moments and N$^3$LO corrections
to their coefficient functions we improve our
previous analysis of the CCFR'97 data for $xF_3$.
The possibility of extracting from the fits of $1/Q^2$-corrections
is analysed using three independent models, including infrared
renormalon one. Theoretical question of applicability
of the renormalon-type inspired large-$\beta_0$ approximation for
estimating corrections to the coefficient functions of odd $xF_3$
and even non-singlet $F_2$ moments are considered. The
comparison with [1/1] Pad\'e estimates is given.
The obtained  NLO and NNLO values of $\alpha_s(M_Z)$ are supporting
the  results
of our less definite previous analysis and are in agreement with
the world average value $\alpha_s(M_Z)\approx 0.118$.
We also present firts
N$^3$LO extraction of $\alpha_s(M_Z)$.
The interplay between higher-order perturbative QCD
corrections and $1/Q^2$-terms is demonstrated.
The results of our studies  are compared with those
obtained recently using the NNLO model of the kernel of the  DGLAP
equation and with the results of the NNLO fits to  CCFR'97 $xF_3$ data,
performed by the  Bernstein polynomial technique.

\vspace*{0.1cm}
\noindent
\\[3mm]
\vfill\eject

\setcounter{page}{1}
\pagestyle{plain}

\section{Introduction}

The series of our previous works \cite{KKPS1}-\cite{KPS2}
(for  a brief summary see the review of  Ref.~\cite{Catani})
was devoted to the QCD analysis of the $xF_3$ data obtained at
the Fermilab Tevatron. In Refs.~\cite{KKPS2}-\cite{KPS2}
we made the consequent steps towards the next-to-next-to-leading
order (NNLO) fits to the $xF_3$ CCFR'97  data \cite{CCFR2}
both without
and with twist-4 corrections taken into account.
In the process of these studies  definite theoretical approximations
were made. This work is devoted to the fixation  of the number
of theoretical ambiguities involved in our previous fits.
Let us recall the
basic steps of these studies.

The $x$-behaviour of $xF_3$  was
reconstructed from the Mellin moments with the number $2\leq  n \leq 10$
with the help of
the Jacobi polynomial technique, developed in Refs.~\cite{PS}
-\cite{Kriv2}.
At the NNLO the perturbative expansion for the coefficient functions
of these moments is explicitly known from analytical
calculations of Refs.~\cite{VZ1}, which were confirmed recently
with the help of another technique \cite{MV}.
However, the NNLO corrections to the anomalous dimensions of the
considered non-singlet (NS) moments, closely related to still analytically
unknown  NNLO
contributions to the kernel of the DGLAP equation \cite{GL}-\cite{D},
were modelled   in the definite approximations only.
These approximations were based on the observation
that for $2\leq n\leq 10$ the next-to-leading (NLO) corrections to the
anomalous dimensions  $\gamma_{NS,F_2}^{(1)}(n)$
of the NS Mellin moments for $F_2$
 do not
numerically differ from the NLO corrections to the anomalous dimensions
$\gamma_{F_3}^{(1)}(n)$ of the moments taken from
the $xF_3$ structure function (SF)  \cite{KKPS1}.
In view of this  it was assumed that this
similarity will be true in  the case of higher-order corrections
to the anomalous dimensions $\gamma_{NS,F_2}^{(n)}$ and $\gamma_{F_3}^{(n)}$
as well,
provided the typical diagrams
with new Casimir structure $d^{abc}d^{abc}$
(which are starting to contribute to $\gamma_{F_3}^{(n)}$ from the NNLO)
are not large. Using this assumption,  we applied  in our analysis
\cite{KKPS2}-\cite{KPS2}
of  the $xF_3$ CCFR'97  data   \cite{CCFR2} the available results for
the NNLO  corrections to
$\gamma_{NS,F_2}^{(2)}(n)$,
which were  known in the cases
of $n=2,4,6,8,10$
due to the  distinguished calculations of Refs.~\cite{LRV,LNRV}.
For the odd moments with $n=3,5,7,9$
the  NNLO corrections to  $\gamma_{NS,F_2}^{(2)}(n)$
were estimated using the procedure of smooth interpolations, previously
proposed in Ref.~\cite{PKK} in the process of first NS NNLO fits
to the  $F_2$ data of the BCDMS collaboration \cite{BCDMS}.

Quite recently the renormalization group quantities
for the $xF_3$ Mellin moments were analytically calculated
at the NNLO level \cite{RV} by the  methods
of Refs.~\cite{Gorishnii:1983su}-\cite{Chetyrkin:1981qh}
, used
in  the case of calculations of the   even Mellin moments
of  $F_2$ \cite{LRV,LNRV}. In Ref.~\cite{RV}
the following information, quite useful for
the fixation of  some  theoretical ambiguities of our previous fits,
was obtained:
\begin{itemize}
\item the NNLO corrections to  $\gamma_{F_3}^{(n)}$
at
$n=3,5,7,9,11,13$;
\item the NNLO corrections to
$\gamma_{NS,F_2}^{(n)}$
at
$n=12$ and $n=14$;
\item the N$^3$LO corrections to the coefficient functions of
odd
Mellin
moments of $xF_3$ with $n=3,5,7,9,11,13$  \footnote{In the case of $n=1$
the  $\alpha_s^3$  contribution to the Gross--Llewellyn-Smith
sum rule was already known \cite{LV}.}.
\end{itemize}

Due to the appearance of  this new important information it became possible to
recast the analysis of Refs.~\cite{KKPS2}-\cite{KPS2} on the new level
of understanding.
First, we are now able to improve
the precision  of the smooth interpolation
procedure for $\gamma_{NS,F_2}^{(n)}$, applied in the  process of
the works of Refs.~\cite{KKPS1}-\cite{KPS2}.
Second, it is now possible
to change the previously-used approximation for NS anomalous dimensions
and take into account
the available  expressions for the  NNLO
corrections to $\gamma_{F_3}(n)$  including terms
typical to this level of perturbation theory, which
are proportional to  $f(d^{abc}d^{abc})/n$
(in our case  we will consider  $f=4$ number of massless active flavours,
while for the $SU_C(3)$-group $(d^{abc}d^{abc})/3=40/9$).
Moreover, since the best way of estimating the   perturbative part of
theoretical uncertainties of the N$^{(i)}$LO fits  is the
incorporation of the  N$^{(i+1)}$LO-terms, the results of Ref.~\cite{RV}
are giving us the chance to perform the approximate N$^3$LO analysis
of $xF_3$ data and compare its  outcome  with the results obtained
within the framework of the N$^3$LO Pad\'e motivated
fits of  Ref.~\cite{KPS2}.

The aims of this work are the following:
\begin{enumerate}
\item To reveal and eliminate  definite theoretical uncertainties
related to the previously-used NNLO approximation $\gamma_{NS,F_2}^{(2)}(n)
\approx \gamma_{F_3}^{(2)}(n)$ at  $2\leq n\leq 10$;
\item To include in the analysis the
NNLO corrections to the coefficient functions $C_{F_3}^{(n)}$
and to the anomalous dimensions $\gamma_{F_3}^{(n)}$
up to $n\leq 13$ (the first results of this program were already
presented in Ref.~\cite{Kataev:2000nc});
\item
To reveal and fix the uncertainties of our previous  NNLO fits of
Refs.~\cite{KKPS2}-\cite{KPS2}, which had the aim
to determine the  NNLO value
of $\alpha_s(M_Z)$ and the parameters of the   $xF_3$ SF model at the
initial scale $Q_0^2$;
\item
To check the
feature of the
interplay between NNLO perturbative QCD corrections
and $1/Q^2$- terms, discovered in
Refs.~\cite{KKPS2,KPS2},  which  leads to the effective
decrease  of the value of the basic free parameter of the
infrared renormalon (IRR) model
for the twist-4 contribution to $xF_3$ \cite{DW}  (for the
discussions of the IRR contributions to characteristics of deep-inelastic
scattering see Refs.~\cite{Dokshitzer:1996qm,Akhoury:1997rt}, while
for recent
reviews of the current general status of the  IRR approach
see Refs.~\cite{Beneke, BB});
\item To comment upon the predictive abilities and special
features of the Pad\'e approximants at the
N$^3$LO level;
\item To study  the applicability of the  renormalon-inspired
large-$\beta_0$ expansion
for estimating higher-order perturbative corrections
to  the coefficient functions of moments of $xF_3$, which
will be used by us. The similar  consideration  of NS moments for $F_2$ SF
~\cite{Mankiewicz:1997gz}
is also updated. The  results will be compared with
those  obtained by means of the  Pad\'e approximation technique;
\item To study the scale-dependence of the obtained values  for
 $\alpha_s$  at the
NLO and beyond;
\item To
reconsider the problem of the  extraction of the $x$-shape of the twist-4
contributions at the LO, NLO, NNLO   using new information about
$\gamma_{F_3}^{(2)}(n)$ and taking into account additional terms in
the perturbative expansion of $C_{F_3}^{(n)}$ at the  N$^3$LO;
\item To compare our results with other phenomenological applications
of the  perturbative QCD
calculations of  Refs.~\cite{VZ1,RV},   which  appeared in
the literature recently and were based on the  application
of DGLAP approach (see Ref.~\cite{Neerven:2001pe}) and
Bernstein polynomial technique
proposed
in Ref.~\cite{Yndurain:1978wz} and used at the NNLO in
Ref.~\cite{ Santiago:2001mh}.
\end{enumerate}

It should be stressed that since in the current work we are interested
in the study of {\it theoretical ambiguities} of the outcome of the fits
on a more solid background than  was done in Ref.~\cite{KPS2},
and in revealing {\it the importance} of the knowledge of {\it exact
expressions} for still uncalculated NNLO corrections to
$\gamma_{F_3}^{(2)}(n)$ for $n$ even,
we
will neglect the systematic experimental  uncertainties of the CCFR'97
$xF_3$ data,
taking
into account statistical ones only. The incorporation to the fits of
systematic error-bars might  shadow the effects of fixing the   theoretical
uncertainties we are struggling for and
might make the process of clarification  of the necessity of getting explicit
numbers for $\gamma_{F_3}^{(2)}(n)$ for even $n$ more complicated.
 At the NLO level the combined
analysis of statistical and systematic experimental  uncertainties of the
$xF_3$
CCFR'97 data was done in  Refs.~\cite{Alekhin:1999df,AK2}
using the machinery of the method
proposed in Ref.~\cite{Alekhin}.

\section{Preliminaries}

For the sake of completeness of the presentation
we will repeat some definitions from Ref.~\cite{KPS2}.

We start from   the Mellin moments
for   $xF_3(x,Q^2)$:
\begin{equation}
M_n^{F_3}(Q^2)=\int_0^1 x^{n-1}F_3(x,Q^2)dx
\end{equation}
where $n=2,3,4,...$. These moments
obey the following renormalization group equation
\begin{equation}
\bigg(\mu\frac{\partial}{\partial\mu}+\beta(A_s)\frac{\partial}
{\partial A_s}
-\gamma_{F_3}^{(n)}(A_s)\bigg)
M_n^{F_3}(Q^2/\mu^2,A_s(\mu^2))=0
\label{rg}
\end{equation}
where $A_s=\alpha_s/(4\pi)$.
The renormalization group functions are defined as
\begin{eqnarray}
\mu\frac{\partial A_s}{\partial\mu}=\beta(A_s)=-2\sum_{i\geq 0}
\beta_i A_s^{i+2}~~\nonumber \\
-\mu\frac{\partial ln
Z_n^{F_3}}{\partial\mu}=\gamma_{F_3}^{(n)}(A_s) =\sum_{i\geq 0}
\gamma_{F_3}^{(i)}(n) A_s^{i+1}
\end{eqnarray}
The NS anomalous dimensions
of $F_2$ will be defined   in the analogous way, namely :
\begin{equation}
-\mu\frac{\partial ln Z_n^{NS,F_2}}{\partial \mu} =
\gamma_{NS,F_2}^{(n)}(A_s)=
\sum_{i\geq 0} \gamma_{NS,F_2}^{(i)}(n) A_s^{i+1}
\end{equation}
where
$ Z_n^{F_3} $ and $Z_n^{NS,F_2}$
are the renormalization constants of the corresponding
NS operators.
In the case of $xF_3$ moments the solution
of the renormalization group
equation is:
\begin{equation}
\frac{ M_{n}^{F_3}(Q^2)}{M_{n}^{F_3}(Q_0^2)}=
exp\bigg[-\int_{A_s(Q_0^2)}^{A_s(Q^2)}
\frac{\gamma_{F_3}^{(n)}(x)}{\beta(x)}dx\bigg]
\frac{C_{F_3}^{(n)}(A_s(Q^2))}
{C_{F_3}^{(n)}(A_s(Q_0^2))}
\label{mom}
\end{equation}
where $M_n^{F_3}(Q_0^2)$ is a phenomenological quantity defined  at  the
initial scale $Q_0^2$ as:
\begin{equation}
M_n^{F_3}(Q_0^2)=\int_0^1 x^{n-2} A(Q_0^2)x^{b(Q_0^2)}(1-x)^{c(Q_0^2)}
(1+\gamma(Q_0^2)x)dx~~~.
\end{equation}
This expression  is
identical to the parametrization used by the CCFR collaboration \cite{CCFR2}.
In the process of our studies we will extract from the fits the
parameters $A(Q_0^2)$, $b(Q_0^2)$, $c(Q_0^2)$ and $\gamma(Q_0^2)$,
together with the parameter $\Lambda_{\overline{MS}}^{(4)}$ and
with the information on the twist-4 terms.

The first moment of $xF_3$ coincides with   the
Gross--Llewellyn Smith sum rule
\begin{equation}
GLS(Q^2)=M_1^{F_3}(Q^2)= \int_0^1 F_3(x,Q^2) dx~~~~.
\end{equation}
At the N$^3$LO
the coefficient function $C_{F_3}^{(n)}$, which enters Eq.~(5),
can be   defined as
\begin{equation}
C_{F_3}^{(n)}(A_s)=1+C^{(1)}_{F_3}(n)A_s+C^{(2)}_{F_3}(n)A_s^2
+C^{(3)}_{F_3}(n)A_s^3,
\end {equation}
while the corresponding  expression  for  the anomalous-dimension term
is :
\begin{equation}
exp\bigg[-\int^{A_s(Q^2)}
\frac{\gamma_{F_3}^{(n)}(x)}{\beta(x)}dx\bigg]=
\big(A_s(Q^2)\big)^{\gamma_{F_3}^{(0)}(n)/2\beta_0}
\times AD(n,A_s)
\end{equation}
where
\begin{equation}
AD(n,A_s)=
[1+p(n)A_s(Q^2)
+q(n)A_s(Q^2)^2+r(n)A_s(Q^2)^3]
\label{an}
\end{equation}
and  $p(n)$, $q(n)$ and $r(n)$ read:
\begin{equation}
p(n)=\frac{1}{2}\bigg(\frac{\gamma_{F_3}^{(1)}(n)}{\beta_1}-
\frac{\gamma_{F_3}^{(0)}(n)}{\beta_0}\bigg)\frac{\beta_1}{\beta_0}
\end{equation}
\begin{equation}
q(n)=\frac{1}{4}\bigg( 2p(n)^2+
\frac{\gamma_{F_3}^{(2)}(n)}{\beta_0}+\gamma_{F_3}^{(0)}(n)
\frac{(\beta_1^2-\beta_2\beta_0)}{\beta_0^3}-\gamma_{F_3}^{(1)}(n)
\frac{\beta_1}{\beta_0^2}\bigg)
\end{equation}
\begin{eqnarray}
r(n)=\frac{1}{6}\bigg(-2p(n)^3+6p(n)q(n)+
\frac{\gamma_{F_3}^{(3)}(n)}{\beta_0}-\frac{\beta_1\gamma_{F_3}^{(2)}(n)}
{\beta_0^2} \\ \nonumber
-\frac{\beta_2\gamma_{F_3}^{(1)}(n)}{\beta_0^2}+
\frac{\beta_1^2\gamma_{F_3}^{(1)}(n)}{\beta_0^3}
-\frac{\beta_1^3\gamma_{F_3}^{(0)}(n)}{\beta_0^4}
-\frac{\beta_3\gamma_{F_3}^{(0)}(n)}{\beta_0^2}+
\frac{2\beta_1\beta_2\gamma_{F_3}^{(0)}(n)}{\beta_0^3}\bigg)~~~.
\end{eqnarray}
The coupling constant $A_s(Q^2)$ can be decomposed
into the inverse powers of
$L=\ln(Q^2/\Lambda_{\overline{MS}}^2)$
as $A_s^{NLO}=A_s^{LO}+\Delta A_s^{NLO}$,
$A_s^{NNLO}=A_s^{NLO}+\Delta A_s^{NNLO}$ and
$A_s^{N^3LO}=A_s^{NNLO}+\Delta A_s^{N^3LO}$, where
\begin{eqnarray}
A_s^{LO}&=&\frac{1}{\beta_0
L} \\ \nonumber
\Delta A_s^{NLO}&=&
-\frac{\beta_1 ln(L)}{\beta_0^3 L^2}
\end{eqnarray}
\begin{equation}
\Delta A_s^{NNLO}=\frac{1}{\beta_0^5 L^3}[\beta_1^2 ln^2 (L)
-\beta_1^2 ln(L) +\beta_2\beta_0-\beta_1^2]
\end{equation}
\begin{eqnarray}
\Delta A_s^{N^3LO}=\frac{1}{\beta_0^7 L^4}[\beta_1^3 (-ln^3 (L)
+\frac{5}{2}ln^2 (L)
+2ln(L)-\frac{1}{2})
\\ \nonumber
-3\beta_0\beta_1\beta_2 ln(L)
+\beta_0^2\frac{\beta_3}{2}]~.
\end{eqnarray}
Notice that in our normalization the  expressions for
$\beta_0$, $\beta_1$, $\beta_2$ and $\beta_3$ have the
following numerical expressions:
\begin{eqnarray}
\beta_0&=&11-0.6667f \nonumber \\
\beta_1&=&102-12.6667f \nonumber \\
\beta_2&=&1428.50-279.611f+6.01852f^2 \nonumber \\
\beta_3&=&29243.0-6946.30f+405.089f^2+1.49931f^3
\end{eqnarray}
where  $\beta_3$ was analytically calculated in
Ref.\cite{RVL}.

\section{Anomalous dimensions and coefficient functions:\\
approximate  vs exact results}

Let us discuss the numerical approximations of higher-order
perturbative corrections to  anomalous dimensions and
coefficient functions,
used in our fits. The analytical expression for the  one-loop term
of NS anomalous dimensions
$\gamma_{NS,F_2}^{(0)}(n)=\gamma_{F_3}^{(0)}(n)=
8/3[4\sum_{j=1}^{n}(1/j)-2/n(n+1)-3]$ is well known.
The NLO corrections to the NS anomalous dimensions were
obtained in Refs.~\cite{1,2} and confirmed by the independent
calculation in Ref.~\cite{3}.
In the cases of both $F_2$ and $xF_3$, the
numerical expressions for the NLO contributions to the anomalous
dimensions are given in Table 1, where
we also present the numerical values of the
NNLO coefficients used  to the NS anomalous dimension
functions of Eq.~(3) and Eq.~(4). They are also normalized to
the case of $f=4$ number of flavours.
The coefficients  $\gamma_{F_3}^{(2)}(n)|_{wts}$ represent the contribution
to $\gamma_{F_3}^{(2)}$ of the terms without  typical structure ({\it wts})
$f(d^{abc}d^{abc})/n=4*40/9$, which is absent in the
expression for $\gamma_{NS,F_2}^{(2)}$ and  appears for the first
time in the anomalous dimensions of $xF_3$ moments at the NNLO.

We  describe now  how the related approximations for
$\gamma_{NS,F_2}^{(2)}(n)$ and $\gamma_{F_3}^{(2)}(n)$
were obtained and what is the connection between the sets
of these numbers. The expressions without round brackets are the
results of explicit analytical calculations of Refs.~\cite{LRV,LNRV,RV}.
The expressions in round brackets are the approximations obtained
with the help  of the smooth interpolation procedure.
To study the possibility of getting  stable dependence of
the  values for
$\Lambda_{\overline{MS}}^{(4)}$ from the change of the initial scale
$Q_0^2$ after  incorporation to the
fits of the NNLO corrections to
$\gamma_{F_3}^{(n)}$
with  $10 < n< 13$,  the interpolation
procedure
was supplemented with the fine tuning of the NNLO corrections
to $\gamma_{F_3}^{(2)}(n)$ with $n=6,8,10$.

\begin{center}
\begin{tabular}{||r|c|c|c|c|c||}
\hline
n  & $\gamma_{NS,F_2}^{(1)}(n)$
&$\gamma_{F_3}^{(1)}(n)$
& $\gamma_{NS,F_2}^{(2)}(n)$
&  $\gamma_{F_3}^{(2)}(n)|_{wts}$
& $\gamma_{F_3}^{(2)}(n)$ \\
\hline
          2&   71.374&      71.241  & 612.0598  & (585) & (631) \\
          3&  100.801&      100.782 & ( 839.8534)&
 836.3440  & 861.6526  \\
          4&  120.145&      120.140 &  1005.823 & (1001.418) &   (1015.368)\\
          5&  134.905&      134.903 &(1134.967) & 1132.727 & 1140.900  \\
          6&  147.003&      147.002 & 1242.0006 & (1241.21)   & (1246.59) \\
          7&  157.332&      157.332 &(1334.865) & 1334.316 & 1338.272 \\
          8&  166.386&      166.386 &  1417.451 & (1416.73) & (1419.783) \\
          9&  174.468&      174.468 &(1491.711) & 1491.124 & 1493.466 \\
         10&  181.781&      181.781 &  1559.005 & (1558.854)& (1560.675)\\
         11&  188.466&      188.466 & (1620.755)& 1620.727 & 1622.283 \\
         12&  194.629&      194.629 & 1678.400  & (1677.696)& (1679.809) \\
         13&  200.350&      200.350 & (1732.640)  & 1731.696 & 1732.809 \\
\hline
\end{tabular}
\end{center}
{{\bf Table 1.} The  numerical expressions for  NLO
and NNLO coefficients of  anomalous dimensions of NS   moments of
$F_2$ and $xF_3$ at $f=4$ number of flavours.}

The numerical values of the obtained fine-tuned numbers
will be given in the next Section.
This procedure
results in better stability of the fitted values of
$\Lambda_{\overline{MS}}^{(4)}$ with respect  to changes of the initial scale
at $Q_0^2\geq 5$ GeV$^2$
(see Table 4  below). The approximation  for $\gamma_{F_3}^{(2)}(2)$ contains
more uncertainties.
At the first stage it was  obtained by extrapolation of the  NNLO coefficients
to anomalous dimensions with $n>2$ without using the explicit number
for $\gamma_{F_3}^{(2)}(n=1)$ (which is zero). We have checked the
reliability of this procedure, considering the set of NLO anomalous dimensions
$\gamma_{F_3}^{(1)}(n)$ which
are explicitly known at any values of $n$. We found that the interpolated
values at even   $n=4,6,...$ are much closer to the real
numbers  than those obtained
with incorporation of  the the zero expression
for the $n=1$ anomalous dimension.
In this case  for $n=2$ we obtained  the extrapolated
value (75.41), which
is over 6$\%$ higher than  the real value 71.24. In the case of application
of the interpolation procedure, which used the $n=1$ result (zero),
we get the value (66.11), which is 6$\%$ smaller than  the real value.
To estimate  the NNLO expression for  $\gamma_{F_3}^{(2)}(2)$ we imposed
the conditions of the reduction found at the NLO and thus
fixed the  value of  $\gamma_{F_3}^{(2)}(2)$ 6$\%$ below
the extrapolated number. Its numerical expression,  which will
be used throughout this work, is quoted in Table 1. The 6$\%$ uncertainty
of $\gamma_{F_3}^{(2)}(2)$ translates to a $1-2$ MeV variation of
$\Lambda_{\overline{MS}}^{(4)}$, which is below
the precision of the NNLO extraction of the value of this parameter.


In view of the doubts in the validity of the interpolation procedure,
expressed in Ref.\cite{SY},
it is of definite interest to desribe the results of
its applications in more detail.
At the first stage
one can compare the interpolated numerical expressions for
$\gamma_{NS,F_2}^{(2)}(n)$ at $n=3,5,7,9,11,13$, obtained with application of
values recently calculated in Ref.~\cite{RV}  for $\gamma_{NS,F_2}^{(2)}(12)$
and $\gamma_{NS,F_2}^{(2)}(14)$ with the explicit identical results for
$\gamma_{F_3}^{(2)}(n)|_{wts}$,
which do not contain the  $d^{abc}d^{abc}$-structure.
The comparison is  presented in  Table 1. The estimates  from
the third column   (839.8534),
(1134.967), (1334.865), (1491.711), (1620.755), (1732.640)
should be compared with the
explicit numbers 836.3440; 1132.727; 1334.316; 1491.124; 1620.727 and
1731.696.
One can see that the qualitative  agreement between these sets of numbers is
rather good. Moreover, we can study the applicability of the interpolation
procedure for simulating explicitly unknown coefficients  of
$\gamma_{NS,F_2}^{(2)}(n)$ obtained with and without application
of $\gamma_{NS,F2}^{(2)}(12)$ and $\gamma_{NS,F2}^{(2)}(14)$ terms.
In fact we found that for $n$=3,5,7,9 the difference
between the ``new'' and ``old'' interpolated  numbers is
not large, namely  2.45;  -0.8; 0.9; -1.8.
It affects  the third significant digit
of the interpolated estimates and  improves their  qualitative agreement
with the results of the
calculations of $\gamma_{F_3}^{(2)}(n)$ at
odd values of $n$ \cite{RV}. However, it is known
that even  a knowledge
of the 3rd significant digit in anomalous dimension terms
is not enough for the precise
reconstruction of SF
from the the NLO results for the
moments at large $n \ge 6$ \cite{Kriv1,Kriv2}.
Therefore, to determine
the unknown even values of $\gamma_{NS,F_3}^{(2)}(n)$
from the known NNLO corrections
to $\gamma_{F_3}^{(2)}(n)$ at $n=3,5,7,9,11,13$
\cite{RV} with more precision
we  supplement   the interpolation
procedure for $\gamma_{F_3}^{(2)}(n)$
by the
fine-tuning of its terms for  $n=6,8,10$.
The results of the  application of the first  approximation
are presented in the last column of Table 1.
Notice that they contain
the scheme-independent new contributions, labelled by the  $d^{abc}d^{abc}$
gauge group structure. The fine-tuned expressions for
$\gamma_{F_3}^{(2)}(n)$ at $n=6,8,10$ will be presented below.

One more verification of the idea of smooth interpolation comes from the
consideration of its application for estimating   NNLO corrections
to the coefficient functions of even Mellin moments of $xF_3$. The estimates
obtained are presented in column 4 of Table 2.
The comparison with the results, given in column 3, which were calculated
from the expression of Ref.~\cite{VZ1}, demonstrate perfect agreement
of these estimates
with the
explicit numbers. We think that in view   of this, one can safely apply
the idea of smooth
interpolation
in order to estimate the N$^3$LO terms to the coefficient
functions of  even Mellin moments of $xF_3$ from those
calculated in Ref.~\cite{RV}  with $n=3,5,7,9,11,13$,
and also taking  into account
the order $O(A_s^3)$ correction to the Gross--Llewellyn Smith
sum rule
($n=1$ moment), obtained in Ref.~\cite{LV}.

The information about the N$^3$LO corrections to the considered
coefficient functions is the important and dominating ingredient
of the N$^3$LO fits to $xF_3$ data we are going to perform. It
should be supplemented with the model for the N$^3$LO corrections $r(n)$ to
the anomalous dimension ($AD$) function of Eq.~(10).

\begin{center}
\begin{tabular}{||r|c|c|c|c|c|c||} \hline
n&
$C_{F_3}^{(1)}(n)$ &  $C_{F_3}^{(2)}(n)$ & $C_{F_3}^{(2)}(n)|_{int}$
& $C_{F_3}^{(3)}(n)|_{int}$ &
$C_{F_3}^{(3)}(n)|_{[1/1]}$ &
$C_{F_3}^{(3)}(n)|_{[0/2]}$ \\ \hline
1&  -4     & -52     &  -52       &   -644.3464  &  -676       &  480      \\
2&  -1.778 & -47.472 & (-46.4295) &  (-1127.454) &  -1267.643  &  174.4079 \\
3&   1.667 & -12.715 &  -12.715   &   -1013.171  &   97.00418  & -47.01328 \\
4&   4.867 &  37.117 & (37.0076)  &  (-410.6652) &   283.0851  &  246.0090 \\
5&   7.748 & 95.4086 &  95.4086   &    584.9453  &   1174.835  &  1013.328 \\
6&  10.351 & 158.2912& (158.4032) &  (1893.575)  &   2420.569  &  2167.903 \\
7&  12.722 & 223.8978& 223.8978   &   3450.468   &   3940.284  &  3637.790 \\
8&  14.900 & 290.8840& (290.8421) &  (5205.389)  &   5678.657  &  5360.371 \\
9&  16.915 & 358.5874&  358.5874  &   7120.985   &   7601.721  &  7291.305 \\
10& 18.791 & 426.4422& (426.5512) &  (9170.207)  &   9677.391  &  9391.308 \\
11& 20.544 & 494.1881&  494.1881  &   11332.82   &   11885.25  & 11633.28  \\
12& 22.201 & 561.5591& (561.2668) &  (13590.97)  &   14204.22  & 13991.80  \\
13& 23.762 & 628.4539&  628.4539  &   15923.91   &   16620.99  & 16449.68  \\
\hline
\end{tabular}
\end{center}
{{\bf Table 2.} The  values for  NLO, NNLO, N$^3$LO
QCD contributions to the coefficient functions,
used in our fits, and the results of  N$^3$LO Pad\'e estimates.}
\vspace{0.5cm}
We  fix them using
the  [1/1] Pad\'e resummation procedure of  the coefficients
of the  $AD$-function
(for the results see Table 3, where the expressions which
come from [0/2] Pad\'e estimates are also presented). It should be mentioned
that the values for $p(n)$ and $q(n)$ are calculated from the numbers
given in Table 1.

\begin{center}
\begin{tabular}{||r|c|c|c|c||} \hline
n&
$p(n)$ & $q(n)$ & $r(n)|_{[1/1]}$ & $r(n)|_{[0/2]}$  \\ \hline
1&  0 &    0    &  0  &    0 \\
 2&   1.6462&  4.8121& 14.0666& 11.3822 \\
 3&   1.9402&  5.5018& 15.6011& 14.0456 \\
 4&   2.0504&  5.8327& 16.5919& 15.2986 \\
 5&   2.1149&  6.2836& 18.6691& 17.1187 \\
 6&   2.1650&  6.7445& 21.0110& 19.0560 \\
 7&   2.2098&  7.1671& 23.2447& 20.8847 \\
 8&   2.2525&  7.6013& 25.6518& 22.8152 \\
 9&   2.2939&  8.0164& 28.0151& 24.7073 \\
10&   2.3344&  8.4353& 30.4804& 26.6614 \\
11&   2.3741&  8.8146& 32.7261& 28.4720 \\
12&   2.4131&  9.1855& 34.9647& 30.2794 \\
13&   2.4512&  9.5620& 37.3002& 32.1491 \\
\hline
\end{tabular}
\end{center}
{{\bf Table 3.} The  values for  NLO and NNLO
QCD contributions to the expanded anomalous dimension terms used
in our fits and the N$^3$LO Pad\'e estimates.}
\vspace{0.5cm}
Several comments should be made concerning the
comparison   of the Pad\'e estimates  technique  of the
N$^3$LO corrections
$C_{F_3}^{(3)}(n)$ with more definite, to our point of view,
results of application
of the interpolation procedure (see Table 2).
One can see that the agreement of [1/1] Pad\'e estimates with the N$^3$LO
coefficients  is good in the case of the Gross--Llewellyn Smith sum rule
(this fact was already known from the estimates of Ref.~\cite{SEK}, which are
close to the results of the scheme-invariant approach of Ref.~\cite{KS}).
In the case of $n=2$ and $n\geq 6$ moments the numbers of columns 5 and 6
of Table 2 are also in satisfactory agreement. Indeed, one should keep in
mind that the difference between the numbers  presented in
columns 5 and 6 of Table 2
should be divided   by the
factor $(4)^3$, which comes from our definition of the  expansion parameter
$A_s=\alpha_s/(4\pi)$. Note that starting from $n\geq 6$ the results
of the application
of [0/2] Pad\'e approximants, which in accordance with the  analysis
of Ref.\cite{Gardi}  reduce scale-dependence uncertainties,
are even closer to the the estimates which are given by
the interpolation procedure
(for the comparison of the outcome  of approximate
N$^3$LO fits to $xF_3$ data, which are based on  [1/1] and
[0/2] Pad\'e approximants, see
 Ref.~\cite{KPS2}, while  the comparison of
the applications of [1/1] and [0/2] Pad\'e approximants
within the quantum mechanic model was analysed in Ref.~\cite{PP}).
For $n=3,4$ the  interpolation method
gives completely different results. The failure of the application of the
Pad\'e estimates approach in these cases might be related   to the irregular
sign structure of the perturbative series under consideration.
A similar problem arises in the case of the analysis of the
perturbative series for QED renormalization group functions (for discussions
see Ref.~\cite{Ellis:1998sb}).

However, in the case of  the perturbative series $AD$ for the expanded
anomalous dimension term
we do not face this problem (see Table 3). In view of the absence of
other ways of fixation of    N$^3$LO coefficients $r(n)$
(the renormalon-inspired large-$\beta_0$ expansion  is definitely
not working for the anomalous dimensions functions
\cite{Mikhailov1, Mikhailov2}) we will use in our fits the [1/1] Pad\'e
estimates for $r(n)$. Note in advance that the application of
[0/2] Pad\'e resummation to $r(n)$ does not influence the outcome
of our approximate N$^3$LO fits with $\alpha_s$ defined by its explicit
N$^3$LO expression (see Eqs.(14)-(16)). It should be stressed that
since in the process of
these fits we will use the explicitly  calculated
coefficient functions of the $xF_3$ Mellin moments~\cite{RV},
the obtained  uncertainties will
be more definite than those estimated in our previous work of
Ref.~\cite{KPS2}.

\section{Results of the fits without twist-4 terms}

In order to perform the concrete fits to $xF_3$ CCFR'97 data and
thus analyze how new theoretical input, described in Sec.3,
affects the results previously obtained in Ref.~\cite{KPS2}, we  apply
the same theoretical method, based on reconstruction of $xF_3$ from
its Mellin moments using the Jacobi polynomial expansion \cite{PS}-
\cite{Kriv2}:
\begin{equation}
\label{eqn:model}
xF_{3}^{N_{max}}(x,Q^2)=
w(\alpha,\beta)(x)
\sum_{n=0}^{N_{max}}
\Theta_n ^{\alpha , \beta}
(x)\sum_{j=0}^{n}c_{j}^{(n)}{(\alpha ,\beta )}
M_{j+2,xF_3}^{TMC}\left ( Q^{2}\right )
+\frac{h(x)}{Q^2}
\end{equation}
where $\Theta_n^{\alpha,\beta}$ are the Jacobi polynomials,
$c_j^{(n)}(\alpha,\beta)$  contain
$\alpha$- and $\beta$-dependent Euler $\Gamma$-functions where
$\alpha,\beta$ are the Jacobi polynomial parameters, fixed by
the minimization of the error in the reconstruction of
the SF, and $w(\alpha,\beta)=x^{\alpha}(1-x)^{\beta}$
is the corresponding weight function with $\alpha=0.7$ and $\beta=3$ chosen
following the detailed analysis of Ref.~\cite{KPS2}.
The contributions of the dynamical twist-4 terms are modelled
by the $Q^2$-independent function $h(x)$.
The kinematical power corrections, namely the
target mass contributions,  are included in the reconstruction formula
of Eq. (\ref{Jacobi}) up to order $O(M_{nucl}^2/Q^2)$-terms:
\begin{equation}
M_{n,xF_3}^{TMC}(Q^2)=M_{n}^{F_3}(Q^2)+\frac{n(n+1)}{n+2}
\frac{M_{nucl.}^2}{Q^2}
M_{n+2}^{F_3}(Q^2)~~~.
\label{TMC}
\end{equation}
Using Eqs.~(\ref{Jacobi}),(\ref{TMC}) one can conclude  that  choosing
$N_{max}=6$, as  was done in the  case of NNLO fits of Refs.~\cite{KKPS2}-
\cite{KPS2}, we  are  taking into account
$2\leq n\leq 10$ Mellin moments in Eq.~(\ref{Jacobi}). As was emphasized
above, definite information on the NNLO QCD corrections to the
renormalization-group functions of $2\leq n\leq 13 $ moments of $xF_3$
is now available. In view of this we can now  increase the number
of $N_{max}$ from $N_{max}=6$ to $N_{max}=9$ and analyze the
changes in the results of the NNLO fits of Ref.~\cite{KPS2} due to
application of additional theoretical information.

We are starting our studies from   the
case when  twist-4
contributions are switched off (namely  $h(x)=0$).
In Table 4 we present the dependence of the extracted values
of $\Lambda_{\overline{MS}}^{(4)}$ from the variations of $N_{max}$ and
of the initial scale $Q_0^2$.
Looking carefully   at Table 4,
one can clearly see that the results of
the  NLO fits are rather stable under  the changes of  $Q_0^2$.
Moreover, they are in agreement with the ones
presented in Table 3 of Ref.~\cite{KPS2}.
However, for $N_{max}=10$ the values of $\chi^2$ are larger than for the case
of $N_{max}=9$. Moreover, we checked that for $Q_0^2$=20 GeV$^2$ and
$N_{max}$=11, despite the fact that the corresponding  value
$\Lambda_{\overline{\rm MS}}^{(4)}=334\pm37$~MeV is comparable with the
one obtained in the  case of choosing  $N_{max}$=10 (see Table 4),
$\chi^2$ continues to increase (we got $\chi^2$=88.9/86).
Therefore, it might be  reasonable to stop at $N_{max}$=9
and thus take into account in Eq.~(18)  13 moments only.

At $N_{max}=6$ new   NNLO results
agree with   the findings
of  Ref.~\cite{KPS2}. They
demonstrate  the same  dependence
of $\Lambda_{\overline{MS}}^{(4)}$ from $Q_0^2$.
Notice that it  has the stability plateau
starting
from $Q_0^2=20$ GeV$^2$ only. However, there is the
important difference between the results of the  NNLO fits of
the current work and the ones of Ref.~\cite{KPS2}.
Indeed, at the  NNLO  Table 4 demonstrates the
widening  of the stability plateau for
$\Lambda_{\overline{MS}}^{(4)}$
to lower $Q_0^2$ values  for
 $7\leq N_{max}\leq 9$ and the minimization  of $\chi^2$-value
at $N_{max}=9$.
This welcome feature is showing  us
the importance  of changing the approximate expressions
for $\gamma_{NS}^{(2)}(n)$ from the
model $\gamma_{F_3}^{(2)}(n)\approx \gamma_{NS,F_2}^{(2)}(n)$ used in
Ref.~\cite{KPS2}
to the new one, which is based on the {\it exact numbers}
for $\gamma_{F_3}^{(2)}(n)$, calculated in Ref.~\cite{RV} {\it
plus} application of the fine-tuning procedure, which in the  case of
$N_{max}=6$ gives $\gamma_{F_3}^{(2)}(6)=1247.4222\pm2.1357$;
for $N_{max}=7$ gives $\gamma_{F_3}^{(2)}(6)=1248.1219\pm1.0359$,
$\gamma_{F_3}^{(2)}(8)=1420.1729\pm4.0854$;
for $N_{max}=8$ gives $\gamma_{F_3}^{(2)}(6)=1248.5610\pm1.2951$,
$\gamma_{F_3}^{(2)}(8)=1419.3301\pm1.5112$;
$\gamma_{F_3}^{(2)}(10)=1561.4299\pm1.4074$;
and for $N_{max}=9$ gives $\gamma_{F_3}^{(2)}(6)=1247.7852\pm0.5091$,
$\gamma_{F_3}^{(2)}(8)=1420.2215\pm0.3337$;
$\gamma_{F_3}^{(2)}(10)=1560.8461\pm 0.2292$.
Within the quoted error bars   these fine-tuned
numbers agree with the
estimates obtained by
smooth  polynomial interpolation and
presented  in the last column of Table 1. The agreement
improves
with the increasing of $N_{max}$. In the case of $N_{max}=9$ the difference is
in the 4th significant digit.

\begin{center}
\begin{tabular}{||c|c|c|c|c|c|c|c|}
\hline
$N_{max}$& $Q_0^2$ ($GeV^2$) & 5 & 8 & 10 & 20 & 50 & 100 \\
\hline
10& LO  & 266$\pm$35 & 265$\pm$36 & 265$\pm$38 & 264$\pm$35 & 264$\pm$36 & 263$\pm$36 \\
  &     &  (113.2)   &  (113.2)   &  (113.2)   &  (113.1)   &  (112.9)    &  (112.6)    \\
\hline
6& NLO & 341$\pm$37 & 341$\pm$37 & 340$\pm$37 & 340$\pm$36  & 339$\pm$37 & 338$\pm$37 \\
  &    &  (85.4)    &  (85.7)    &   (85.7)   &  (85.7)     &  (85.4)    &  (85.1)    \\
7& NLO & 342$\pm$38 & 341$\pm$38 & 341$\pm$37 & 340$\pm$37  & 339$\pm$37 & 338$\pm$37 \\
  &    &  (87.1)    &  (87.3)    &   (87.4)   &  (87.4)     &  (87.2)    &  (86.9)    \\
8& NLO & 346$\pm$38 & 344$\pm$38 & 344$\pm$38 & 343$\pm$37  & 341$\pm$38 & 341$\pm$38 \\
  &    &  (86.0)    &  (86.4)    &   (86.5)   &  (86.5)     &  (86.2)    &  (85.9)    \\
9& NLO & 349$\pm$42 & 347$\pm$37 & 347$\pm$37 & 345$\pm$38  & 344$\pm$37 & 343$\pm$38 \\
  &    &  (84.2)    &  (84.8)    &   (85.0)   &  (85.1)     &  (84.8)    &  (84.5)    \\
10& NLO& 341$\pm$35 & 340$\pm$33 & 339$\pm$34 & 338$\pm$40  & 337$\pm$37 &
336$\pm$35 \\
  &    &  (87.1)    &  (87.4)    &   (87.5)   &  (87.6)     &  (87.5)    &  (87.3)    \\
\hline
6& NNLO & 297$\pm$30 & 314$\pm$34 & 320$\pm$34 & 327$\pm$36 & 327$\pm$35 &
326$\pm$35\\
 &      &  (77.9)    &   (76.3)   &  (76.2)    &  (76.9)    &  (78.5)   &
(79.5) \\

7& NNLO & 326$\pm$34 & 327$\pm$35 & 327$\pm$35 & 326$\pm$36 & 327$\pm$36 &
328$\pm$35\\
 &      &  (75.9)    &   (76.7)   &  (77.1)    &  (78.1)    &  (78.8)   &
(78.7) \\

8& NNLO & 334$\pm$35 & 334$\pm$35 & 333$\pm$35 & 331$\pm$35 & 328$\pm$35 &
328$\pm$35\\
 &      &  (74.3)    &   (75.7)   &  (76.2)    &  (77.4)    &  (78.3)   &
(78.5) \\

9& NNLO & 330$\pm$33 & 332$\pm$35 & 333$\pm$34 & 331$\pm$37 & 330$\pm$35 &
329$\pm$35\\
 &      &  (72.4)    &   (73.6)   &  (74.7)    &  (75.8)    &  (76.7)   &
(77.8) \\
\hline


6& N$^3$LO  & 303$\pm$29 & 317$\pm$31 & 321$\pm$32 & 325$\pm$33 & 325$\pm$
33 & 324$\pm$33 \\
 &          &  (76.4)    &  (75.6)    &  (75.7)    &  (76.6)    & (78.0) &
(78.7) \\

7& N$^3$LO  & 328$\pm$32 & 326$\pm$33 & 325$\pm$33 & 322$\pm$33 & 324$\pm$
33 & 324$\pm$33 \\
 &          &  (76.2)    &  (77.0)    &  (77.3)    &  (78.2)    & (78.5) &
(78.2) \\

8& N$^3$LO  & 334$\pm$33 & 329$\pm$33 & 327$\pm$34 & 324$\pm$34 & 323$\pm$
34 & 324$\pm$34 \\
 &          &  (74.8)    &  (76.2)    &  (76.6)    &  (77.4)    & (77.3) &
(77.2) \\

9& N$^3$LO  & 330$\pm$31 & 329$\pm$34 & 329$\pm$32 & 325$\pm$33 & 325$\pm$
32 & 325$\pm$33 \\
 &          &  (73.3)    &  (74.6)    &  (75.7)    &  (76.4)    & (76.7) &
(76.8) \\
\hline
\end{tabular}
\end{center}
{{\bf Table 4.} The $Q_0^2$ and $N_{max}$
dependence of $\Lambda_{\overline{MS}}^{(4)}$
[MeV].  The values of $\chi^2$ are presented in
parenthesis.}

However, the improved $Q_0^2$-independent NNLO values of
$\Lambda_{\overline{MS}}^{(4)}$ do not differ significantly
from the results of Ref.~\cite{KPS2} (the difference  of over 4--5 MeV
is about 7 times smaller than the existing statistical error).
A similar feature  reveals itself  in the process of
the   N$^3$LO fits,
which are based in part on application of the {\it exact numbers} for
the $\alpha_s^3$-corrections to the coefficient
functions of odd Mellin moments, calculated in  Ref.~\cite{RV}, and
were performed using  the N$^3$LO approximation of $\alpha_s$.
It should be stressed that   the inclusion in the fits of
these numbers at $N_{max}=6$ does not lead to the detectable
difference of the new results from the ones obtained in Ref.~\cite{KPS2}
with the help of the Pad\'e approximation method.
The essential
advantage of the new considerations  is that we are able to reach $N_{max}=9$
and observe perfect stability of both NNLO and new N$^3$LO
values  for $\Lambda_{\overline{\rm MS}}^{(4)}$ to the  variation of
$Q_0^2$ after slight modification of
    $\gamma_{F_3}^{(2)}(6)$,
$\gamma_{F_3}^{(2)}(8)$, and
$\gamma_{F_3}^{(2)}(10)$ (which arise  from the  application of  fine-tuning
procedure at the N$^3$LO),  and can  determine      from these numbers the
Pad\'e approximations for
$r(n)|_{[1/1]}$ at $n=6,8,10$. It should be mentioned  that
the obtained estimates  for the fitted three terms of $\gamma_{F_3}^{(2)}(n)$
are in agreement with the estimates  presented above at the level
of 4 significant digits.
Note also, that if  we  use  in the fits    [0/2] Pad\'e approximants
for modelling  the N$^3$LO  correction $r(n)$,
we  get the values of
$\Lambda_{\overline{MS}}^{(4)}$, which are rather close to those
obtained   in
the case of application of [1/1] Pad\'e approximants.

Comparing now the results of the NNLO and approximate N$^3$LO fits
we conclude that for $7\leq N_{max}\leq 9$ the difference
between the obatined values of $\Lambda_{\overline{MS}}^{(4)}$
is rather small and almost disappears for  $Q_0^2=5$ GeV$^2$.
Thus we can make the conclusion that we observe the minimization
of theoretical uncertainties and, probably, the saturation of the
predictive power of the corresponding perturbative series at the 4-loop level.
A similar feature was discovered  in the process of
calculations of the perturbative corrections to the correlator
of scalar quark currents in the large-$f$ approximation \cite{BKM}.
Therefore, we think that the perturbation theory
approximants for  $xF_3$ moments can  be safely truncated
at  one more step beyond the NNLO. Higher-order calculations
might  manifest the signal for  asymptotic divergence of the
related perturbative QCD predictions and as  a result, the
increase of the value of $\chi^2$.

\begin{center}
\begin{tabular}{||c|c|c|c|c|c|c||}
\hline
order/$N_{max}$ & $Q_0^2$&  A  & b & c &
$\gamma$ &
$\chi^2$/nop \\ \hline
LO/9 &
5 GeV$^2$
  & 5.13$\pm$0.21 & 0.72$\pm$0.01 & 3.87$\pm$0.06 & 1.42$\pm$0.20
 & 113.2/86  \\
& 10 GeV$^2$
   & 5.07$\pm$0.22 & 0.70$\pm$0.01 & 3.97$\pm$0.08 & 1.16$\pm$0.25
 & 113.2/86 \\
& 20 GeV$^2$
  & 4.98$\pm$0.47 & 0.68$\pm$0.03 & 4.05$\pm$0.10 & 0.96$\pm$0.41
 & 113.1/86  \\
& 100 GeV$^2$
  & 4.73$\pm$0.36 & 0.64$\pm$0.02 & 4.19$\pm$0.09 & 0.62$\pm$0.30
 & 112.6/86  \\ \hline
NLO/9 &
5 GeV$^2$
  &  4.05$\pm$0.20 & 0.65$\pm$0.02 & 3.71$\pm$0.06 & 1.93$\pm$0.16
 & 87.1/86  \\
&10 GeV$^2$
   & 4.48$\pm$0.21 & 0.66$\pm$0.02 & 3.85$\pm$0.05 & 1.32$\pm$0.15
 & 87.5/86  \\
&20 GeV$^2$ & 4.48$\pm$0.21
    & 0.65$\pm$0.01 & 3.96$\pm$0.07 & 0.95$\pm$0.15
& 87.6/86  \\
&100 GeV$^2$
   & 4.73$\pm$0.38 & 0.62$\pm$0.02 & 4.12$\pm$0.12 & 0.46$\pm$0.34
 & 87.3/86  \\
\hline
NNLO/6 & 5 GeV$^2$
   &  4.25$\pm 0.38$ &  0.66$\pm$0.03 &
3.56$\pm$0.07 &  1.33$\pm$0.33
 &  78.4/86  \\

{\bf NNLO/9} &
   & {\bf 3.73$\pm$0.68} & {\bf 0.63$\pm$0.05} & {\bf
3.52$\pm$0.08} & {\bf 1.69$\pm$0.68}
 & {\bf 72.4/86}  \\

NNLO /6 &10 GeV$^2$
   &  4.50$\pm$0.36 &  0.65$\pm$0.03 &
3.73$\pm$0.07 &
 1.05$\pm$0.31 &  76.3/86  \\
{\bf NNLO/9}  &
   & {\bf 4.21$\pm$0.35} & {\bf 0.63$\pm$0.03} & {\bf
3.73$\pm$0.07} &
{\bf 1.22$\pm$0.31} & {\bf 74.2/86}  \\

NNLO/6 &20 GeV$^2$
   &  4.70$\pm$0.34 &  0.65$\pm$0.03 &
3.88$\pm$0.08 &  0.80$\pm$0.30
 &  77.0/86  \\
{\bf NNLO/9} &
   & {\bf 4.49$\pm$0.25} & {\bf 0.63$\pm$0.02} & {\bf
3.89$\pm$0.06} & {\bf 0.93$\pm$0.20}
 & {\bf 75.8/86}  \\

NNLO/6 &100 GeV$^2$
   &  4.91$\pm$0.28 &  0.63$\pm$0.02 &
4.11$\pm$0.10 &  0.53$\pm$0.27
 &  80.0/86  \\
{\bf NNLO/9} &
  & {\bf 4.74$\pm$0.32} & {\bf 0.61$\pm$0.02} & {\bf
4.14$\pm$0.09} & {\bf 0.46$\pm$0.27}
 & {\bf 77.8/86}  \\

\hline
N$^3$LO/9  &
5 GeV$^2$
   & 4.16$\pm$0.28 & 0.65$\pm$0.02 & 3.31$\pm$0.09 & 0.91$\pm$0.21
 & 73.3/86  \\
N$^3$LO/9  &10 GeV$^2$
   & 4.49$\pm$0.41 & 0.65$\pm$0.03 & 3.61$\pm$0.08
& 0.81$\pm$0.32
 & 75.1/86  \\
N$^3$LO/9 &20 GeV$^2$
   & 4.64$\pm$0.72 & 0.64$\pm$0.05 & 3.83$\pm$0.15 & 0.73$\pm$0.60
 &  76.4/86 \\
N$^3$LO/9  &100 GeV$^2$
   &  4.77$\pm$0.30 & 0.61$\pm$0.02   & 4.15$\pm$0.09 &
0.47$\pm$0.26 & 77.6/86   \\
\hline
\end{tabular}
\end{center}
{{\bf Table 5.} The determined
values of the parameters $A,b,c,\gamma$ of the model
for $xF_3$ and their  comparison with the values obtained
in  Ref.~\cite{KPS1}. The new ones, related to the NNLO, are marked
by bold type.}

Another consequence  of our new improved analysis corresponds to the
determination of $Q_0^2$-dependence of  the
parameters $A$, $b$, $c$, and $\gamma$. It is  presented in Table 5
and is  compared with the previous extraction of their  $Q_0^2$-dependence
given in Ref.~\cite{KPS1}. The LO and NLO
numbers  are the same, while at the NNLO the new results
marked out  by bold type, which correspond to  $N_{max}=9$,
are compared with the
previous NNLO ones \cite{KPS1}, obtained  at $N_{max}=6$ using the
calculations of $\gamma_{NS,F_2}^{(2)}(n)$-terms
\cite{LRV,LNRV} available at this time.
One can see the noticeable decrease in $\chi^2$ at the NNLO.
In the LO the obtained $Q_0^2$-dependence  of $c$ is in agreement
with the  shape of variation of this parameter,
predicted in Ref.~\cite{Korch} and confirmed recently in Ref.~\cite{Albino:2000cp}.

Even more interesting to study our NNLO results for the
parameter $b(Q_0^2)$. First, it is almost
$Q_0^2$-independent within statistical errors.
This fact is in agreement with theoretical
demonstration of its $Q_0^2$-independence, presented
in Ref.\cite{Kotikov} using DGLAP equation.
Another result was obtained recently in
Ref.~\cite{Ermolaev:2001sg} in all-orders of
1-loop expression for $\alpha_s$, using in part
the approach, developed in  Ref.~\cite{Kirschner:1983di}.
It should be stressed that the estimate  $\omega^{-}=0.4$
in the expression $ F_{NS}=(1/x)^{\omega^{-}}\big(\frac{Q^2}{\mu^2}\big)^
{\omega^{-}/2}$, obtained
in Ref.~\cite{Ermolaev:2001sg} for $\mu\approx 5.5~{\rm GeV}$,
$\Lambda_{QCD}=0.1$ and $f=3$    is in good
numbers  for $1-b$
especially in the region $Q^2>\mu^2=30~{\rm GeV}^2$, which is not
considerebly affected by the transformation from
$\Lambda^{(3)}\approx 0.1~{\rm GeV}$, used in Ref.~\cite{Ermolaev:2001sg},
to $\Lambda^{(3)}_{\overline{\rm MS}}\approx 0.4~{\rm GeV}$, as advocated
by us.

\section{Incorporation of the twist-4 terms}
\subsection{Infared renormalon model parameterization}
The next stage in modification of
QCD theoreretical approximations
is the inclusion of the higher-twist terms
in the expression for the structure
functions. At the first stage we will
rely on the prediction of the IRR approach \cite{DW}
and model the twist-4 contribution to $h(x)$ in Eq.(\ref{eqn:model}) as
\begin{equation}
\frac{h(x)}{Q^2}=w(\alpha,\beta)\sum_{n=0}^{N_{max}}\Theta_n^{\alpha,\beta}(x)
\sum_{j=0}^{n}c_j^{(n)}(\alpha,\beta)M_{j+2,xF_3}^{IRR}(Q^2)
\end{equation}
where
\begin{equation}
M_{n,xF_3}^{IRR}(Q^2)= \tilde{C}(n)M_n^{F_3}(Q^2)
\frac{A_2^{'}}{Q^2} {\rm~ with}~ \tilde{C}(n)=-n-4+2/(n+1)+4/(n+2)
+4S_1(n) .
\end{equation}
The results of the new improved fits to CCFR'97 data for $xF_3$ with
the twist-4 term taken into account  through
Eq.~(21)
are presented in Table 6.
\begin{center}
\begin{tabular}{||c|c|c|c|c||}
\hline
order/$N_{max}$ &  $Q_0^2=$&
5~Gev$^2$& 20~GeV$^2$  & 100~GeV$^2$ \\ \hline

LO/6 & $\Lambda_{\overline{MS}}^{(4)}$ & 433$\pm$54 & 431$\pm$36 & 429$\pm$35\\
          &$\chi^2$/nep & 81.2/86 & 81.2/86 & 80.6 \\
          &$A_2^{'}$ & $-$0.331$\pm$0.057 & $-$0.330$\pm$0.059 & $-$0.328$\pm$0.058 \\
LO/9 & $\Lambda_{\overline{MS}}^{(4)}$ & 447$\pm$54 & 443$\pm$54 &439$\pm$56\\
       &$\chi^2$/nep & 79.8/86 & 80.1/86 & 79.6/86 \\
       &$A_2^{'}$ & $-$0.340$\pm$0.059 & $-$0.337$\pm$0.059 & $-$0.335$\pm$0.059 \\
\hline

NLO/6 & $\Lambda_{\overline{MS}}^{(4)}$ & 370$\pm$38 & 369$\pm$41 & 367$\pm$38\\
          &$\chi^2$/nep & 80.2/86 & 80.4/86 & 79.9/86 \\
          &$A_2^{'}$ & $-$0.121$\pm$0.052 & $-$0.121$\pm$0.053 & $-$0.120$\pm$0.052 \\
NLO/9 & $\Lambda_{\overline{MS}}^{(4)}$ & 379$\pm$41 & 376$\pm$39 &374$\pm$42\\
       &$\chi^2$/nep & 78.6/86 & 79.5/86 & 79.0 \\
       &$A_2^{'}$ & $-$0.125$\pm$0.053 & $-$0.125$\pm$0.053 & $-$0.124$\pm$0.053 \\

\hline

NNLO/6 & $\Lambda_{\overline{MS}}^{(4)}$ & 297$\pm$30 & 328$\pm$36 &
328$\pm$35\\
          &$\chi^2$/nep & 77.9/86 & 76.8/86 & 79.5/86 \\
          &$A_2^{'}$ & $-$0.007$\pm$0.051 & $-$0.017$\pm$0.051 & $-$0.015$\pm$0.051 \\
NNLO/7 & $\Lambda_{\overline{MS}}^{(4)}$ & 327$\pm$34 & 327$\pm$35 &328$\pm$36
\\
       &$\chi^2$/nep & 75.8/86 & 78.1/86 & 78.6/86 \\
       &$A_2^{'}$ & $-$0.011$\pm$0.051 & $-$0.013$\pm$0.051 & $-$0.015$\pm$0.051 \\
NNLO/8 & $\Lambda_{\overline{MS}}^{(4)}$ & 335$\pm$37 & 330$\pm$36 &329$\pm$36
\\
       &$\chi^2$/nep & 73.8/86 & 77.0/86 & 77.5/86 \\
       &$A_2^{'}$ & $-$0.012$\pm$0.051 & $-$0.013$\pm$0.051 & $-$0.015$\pm$0.051 \\
NNLO/9 & $\Lambda_{\overline{MS}}^{(4)}$ & 331$\pm$33 & 332$\pm$35 &331$\pm$35
\\
       &$\chi^2$/nep & 73.1/86 & 75.7/86 & 76.9/86 \\
       &$A_2^{'}$ & $-$0.013$\pm$0.051 & $-$0.015$\pm$0.051 & $-$0.016$\pm$0.051 \\
\hline
N$^3$LO/6 & $\Lambda_{\overline{MS}}^{(4)}$ & 305$\pm$29 & 327$\pm$34 &326$
\pm$34
\\
       &$\chi^2$/nep & 76.0/86 & 76.2/86 & 78.5/86 \\
       &$A_2^{'}$ & 0.036$\pm$0.051 & 0.033$\pm$0.052 & 0.029$\pm$0.052 \\
N$^3$LO/7 & $\Lambda_{\overline{MS}}^{(4)}$ & 331$\pm$33 & 325$\pm$34 &326$\pm$34
\\
       &$\chi^2$/nep & 75.6/86 & 77.7/86 & 77.8/86 \\
       &$A_2^{'}$ & 0.040$\pm$0.052 & 0.036$\pm$0.052 & 0.035$\pm$0.052 \\
N$^3$LO/8 & $\Lambda_{\overline{MS}}^{(4)}$ & 337$\pm$34 & 326$\pm$34 &326$\pm$34
\\
       &$\chi^2$/nep & 74.1/86 & 76.9/86 & 76.7/86 \\
       &$A_2^{'}$ & 0.040$\pm$0.052 & 0.036$\pm$0.052 & 0.035$\pm$0.052 \\

N$^3$LO/9 & $\Lambda_{\overline{MS}}^{(4)}$ & 333$\pm$34 & 328$\pm$33 &328$\pm$38
\\
       &$\chi^2$/nep & 73.8/86 & 75.9/86 & 76.4/86 \\
       &$A_2^{'}$ & 0.038$\pm$0.052 & 0.035$\pm$0.052 & 0.034$\pm$0.052 \\
\hline
\end{tabular}
\end{center}
{{\bf Table 6.} The results of the fits to  the CCFR'97   $xF_3$ data
with HT terms modelled through the IRR model.
$A_2{'}$   is the additional
 parameter of the fit. The cases of  different $Q_0^2$ and $N_{max}$
 are considered.}

\newpage

Looking carefully at Table 4 we arrive at the following new  conclusions:
\begin{itemize}
\item The $\chi^2$-value decreases from LO up to NNLO and at
N$^3$LO level it almost coincides with the one obtained at the NNLO.
Moreover, $\chi^2$  decreases with the  increasing of  $N_{max}$ and
distinguishes
the fits with   $N_{max}=9$. This is the welcome feature
of including in the fitting procedure more detailed information on
the perturbative theory contributions both
to coefficient functions and anomalous dimensions
of $xF_3$ moments, and in particular explicitly-calculated
 three-loop coefficients
$\gamma_{F_3}^{(2)}(n)$ at $n=3,5,7,9,11,13$ \cite{RV},
supplemented by us with application of
the interpolation procedure for even $n$ {\it plus}
fine-tuning of the terms $\gamma_{F_3}^{(2)}(6)$, $\gamma_{F_3}^{(2)}(8)$
and $\gamma_{F_3}^{(2)}(10)$.
\item At the N$^3$LO, $\chi^2$ is smaller than the one obtained
in the process of pure Pad\'e-motivated fits of Ref.~\cite{KPS2}.
This feature is related to the fact that the explicit expressions
for N$^3$LO corrections for odd  $xF_3$ moments \cite{RV} are now
taken into account.
\item For $N_{max}=9$ the values of $\Lambda_{\overline{\rm MS}}^{(4)}$
and the IRR model free parameter $A_2^{'}$ are rather stable
to variation of $Q_0^2$ not only at the LO, NLO but at the NNLO
and N$^3$LO as well. The last property  gives  favour to   our new
results in comparison with  the ones obtained in Ref.~\cite{KPS2} in the case
of $N_{max}=6$ and $Q_0^2=20~{\rm GeV}^2$,
taking into account a more approximate model for
$\gamma_{F_3}^{(2)}(n)$ and Pad\'e approximations for
$C_{F_3}^{(3)}(n)$ at $n\leq 10$.
\item At the scale $Q_0^2$=20 GeV$^2$  the obviously
visible difference with
the findings  of Ref.~\cite{KPS2} is related to the switch
from the
Pad\'e approximant estimates of  N$^3$LO contributions
to  coefficient functions
to the  expressions  for $C^{(3)}_{F_3}|_{int}$, presented in Table 2,
and obtained from the  calculations
of $C^{(3)}_{F_3}(n)$ at  $n=1$ \cite{LV} and $n=3,5,7,9,11,13$
~\cite{RV}.
The positive outcome of this change is  the shift
of $\Lambda_{\overline{MS}}^{(4)}$ from 340$\pm$ 37 MeV to
the the values, given in Table 6 ( which are  less different from the results
of the   NNLO fits)
and the minimization of the  value of $\chi^2$.
\item The LO and NLO fits seem to support the IRR model for
the twist-4 terms by the foundation of negative values of
$A_2^{'}$, which are different from zero within the presented error-bars
and are  in agreement with the results of the previous
similar fits of Refs.~\cite{KKPS2,KPS2} and with the
one, obtained in Ref.~\cite{Alekhin:1999df} using the NLO DGLAP analysis
of the same set of CCFR'97 data.
\item At the NNLO the central value of $A_2^{'}$ is also negative,
but has large error bars.
Moreover, the inclusion of the
N$^3$LO corrections clearly demonstrates the effective minimization
of the free parameter of the IRR model, which
becomes positive, but has  statistical uncertainties twice
as large as   the central value.
Thus, we may conclude, that
at this level the interplay bewteen high-order perturbative
corrections  and the model for  twist-4
contributions,
discussed from various points of view in Refs.\cite{PP2}--
\cite{Dokshitzer:1999ai}, is manifesting itself.
\end{itemize}

\subsection{IRR approach and naive non-Abelianization}

Perturbative expansion of the IRR model is usually understood
within the framework of the large-$\beta_0$ expansion, where
$\beta_0$ is the first coefficient of the QCD $\beta$-function.
The approximations for  coefficient functions
(but not anomalous dimensions)
obtained within this model can be  compared with
the explicit expression for the quantities under consideration, calculated
in the $\overline{\rm MS}$--scheme.
As a rule, the qualitative success of the estimating power
of the large-$\beta_0$ expansion is rather satisfactory
(see e.g. \cite{Beneke,BKM}).
Let us study the application
of this approach to the coefficient functions of
odd Mellin moments of $xF_3$ used in our work.

They can be presented
in the  following numerical
form (see Refs.~\cite{LV,RV}):
\begin{eqnarray}
C_{F_3}^{(1)}&=&1-4A_s+A_s^2(-73.333+5.333f)+A_s^3(-2652.154+513.310f-11.358f^2) \\ \nonumber
C_{F_3}^{(3)}&=&1+1.667A_s+A_s^2(14.254-6.742f)+A_s^3(-839.764-45.099f+1.748f^2) \\ \nonumber
C_{F_3}^{(5)}&=&1+7.748A_s+A_s^2(173.001-19.398f)+A_s^3(4341.081-961.276f+
22.241f^2) \\ \nonumber
C_{F_3}^{(7)}&=&1+12.722A_s+A_s^2(345.991-30.523f)+A_s^3(11119.001-1960.237f+43.104f^2) \\ \nonumber
C_{F_3}^{(9)}&=&1+16.915A_s+A_s^2(520.006-40.355f)+A_s^3(18771.996-2975.924f
+63.171f^2) \\ \nonumber
C_{F_3}^{(11)}&=&1+20.548A_s+A_s^2(690.872-49.171f)+A_s^3(26941.480-3984.412f+
82.246f^2)
 \\ \nonumber
C_{F_3}^{(13)}&=&1+23.762A_s+A_s^2(857.178-57.181f)+A_s^3(35426.829-4976.081f+100.351f^2)
\end{eqnarray}

Note that for the reasons discussed below,
the order $O(A_s^3)$-corrections to Eq.~(22)
are presented without typical structures ($wts$), proportional
to
$d^{abc}d^{abc}$-terms.

The procedure of large $\beta_0$-expansion is formulated
in our notations
in the following
way: one should extract from the explicit  expressions
for perturbative coefficients, calculated in the ${\overline{\rm MS}}$-scheme,
the leading terms
in the number of flavours $f$  and  then make the substitution
$f\rightarrow-6\tilde{\beta}_0$, where $\tilde{\beta}_0=\beta_0/4$.
This approximation is known in the literature as the ``naive
non-Abelianization'' (NNA)  procedure \cite{GB}. Note that it
does not simulate $d^{abc}d^{abc}$-terms.
Using this pattern  we present below
the coefficient functions of Eq.~(22) in the NNA form:
\begin{eqnarray}
C_{F_3}^{(1)}&=&1-4A_s+A_s^2(-6\times5.333\tilde{\beta}_0)+
A_s^3(-36\times11.358\tilde{\beta}_0^2) \\ \nonumber
C_{F_3}^{(3)}&=&1+1.667A_s+A_s^2(6\times6.742\tilde{\beta}_0)+
A_s^3(36\times1.748f\tilde{\beta}_0^2) \\ \nonumber
C_{F_3}^{(5)}&=&1+7.748A_s+A_s^2(6\times19.398
\tilde{\beta}_0)+A_s^3(36\times
22.241\tilde{\beta}_0^2) \\ \nonumber
C_{F_3}^{(7)}&=&1+12.722A_s+A_s^2(6\times30.523
\tilde{\beta}_0)+A_s^3(36\times43.104\tilde{\beta}_0^2) \\ \nonumber
C_{F_3}^{(9)}&=&1+16.915A_s+A_s^2(6\times40.355\tilde{\beta}_0)+A_s^3(
36*63.171\tilde{\beta}_0^2) \\ \nonumber
C_{F_3}^{(11)}&=&1+20.548A_s+A_s^2(6\times49.171
\tilde{\beta}_0)+A_s^3(36\times
82.246\tilde{\beta}_0^2)
 \\ \nonumber
C_{F_3}^{(13)}&=&1+23.762A_s+A_s^2(6\times57.181\tilde{\beta}_0)+A_s^3
(36\times100.351\tilde{\beta}_0^2)
\end{eqnarray}

The obtained  NNLO and N$^3$LO  corrections
should be compared with the corresponding  ones
of Eqs.~(24).
The $f$-dependence
of the ratios $R_{F_3,NNA}^{(2)}(n)=C_{F_3}^{(2)}(n)_{NNA}/C_{F_3}^{(2)}(n)$
and $R_{F_3,NNA}^{(3)}(n)=C_{F_3}^{(3)}(n)_{NNA}/C_{F_3}^{(3)}(n)|_{wts}$,
which follow from the comparison of the related expressions of Eqs.~(22,23),
is presented in Table 7 and Table 8, where the
$f$-dependence of the ratio $R_{F_3}^{(3)}(n)|_{[Pade]}=
C_{F_3}^{(3)}(n)|_{[1/1]}/C_{F_3}^{(3)}(n)|_{wts}$ is also given in
round brackets.

\begin{center}
\begin{tabular}{||r|c|c|c|c||} \hline
$n$ & $f=3$ & $f=4$ & $f=5$
 \\ \hline
1 & 1.25 &   1.28    &  1.31 \\
3 & -15.24&  -6.63 & -3.98 \\
5 & 2.34&  2.61 & 3.07 \\
7 &  1.43 &  1.42 & 1.41 \\
9 &  1.36 &  1.41 & 1.46 \\
11& 1.22 &  1.24 & 1.27 \\
13& 1.13 &  1.14 & 1.15 \\
\hline
\end{tabular}
\end{center}
{{\bf Table 7.} The $f$-dependence of the  ratios
$R_{F_3,NNA}^{(2)}(n)$. }

\begin{center}
\begin{tabular}{||r|c|c|c|c||} \hline
$n$ & $f=3$ & $f=4$ & $f=5$
 \\ \hline
1& 1.61 ({\bf 0.68}) &   2.01 ({\bf 0.87})    &  2.99 ({\bf 1.48}) \\
3& -0.32 ({\bf -0.002})&  -0.26 ({\bf -0.1})  & -0.21 ({\bf -0.22})  \\
5 & 2.46 ({\bf 1.03})&  4.15 ({\bf 1.38})& 41.32 ({\bf 8.22})\\
7&  1.4 ({\bf 0.90})&  1.7 ({\bf 0.99})& 2.4 ({\bf 1.22}) \\
9&  1.1 ({\bf 0.90})&  1.25 ({\bf 0.96}) & 1.56 ({\bf 1.09}) \\
11& 0.95 ({\bf 0.91}) &  1.04 ({\bf 0.96}) & 1.20 ({\bf 1.06}) \\
13& 0.85 ({\bf 0.92})  &  0.95 ({\bf 0.97})  & 1.02 ({\bf 1.05}) \\
\hline
\end{tabular}
\end{center}
{{\bf Table 8.} The $f$-dependence of the ratios
$R_{F_3,NNA}^{(3)}(n)$ and $R_{F_3}^{(3)}(n)|_{[Pade]}$
 (in round brackets). }


One can see that the NNA approach is working reasonably
well in the case
of the Gross-Llewellyn Smith sum rule ($n=1$ moment). This fact
was already observed  in the review of Ref.~\cite{Beneke}.
It also gives satisfactory estimates both at the NNLO and N$^3$LO
in the case  of odd values of $n$ with  $n\geq 7$,
but does not work for $n=3$ (where even the wrong
sign is obtained) and for $n=5$, where at the N$^3$LO the
subleading in $f$ (and thus $\beta_0$) term is larger than the leading
$\beta_0^2$-contribution.
Notice that in the case of even NS moments of
$F_2$  the situation was the same: the NNA approximation was predicting the
correct sign starting from the   $n=4$ moment and was giving
qualitatively good
estimates in the cases of $n=6,8$ moments~\cite{Mankiewicz:1997gz}.
Armed by the new information about explicit behaviour of the
NS moments for $F_2$ with $n=10$ \cite{LNRV} and $n=12,14$
\cite{RV}, we extend the considerations of Ref.~\cite{Mankiewicz:1997gz}
to the case of higher moments, omitting
$f\sum_{f=1}^{f} e_f$-contribution to the
order O($A_s^3$)-corrections of the NS moments of $F_2$.
Taking into account this approximation we present  the
explicit
expressions for the coefficient functions of the NS moments
of $F_2$ in the following numerical form:
\begin{eqnarray}
C_{F_2,NS}^{(2)}&=&1+0.444A_s+A_s^2(17.694-5.333f)+
A_s^3(442.741-165.197f+6.030f^2) \\ \nonumber
C_{F_2,NS}^{(4)}&=&1+6.607A_s+A_s^2(141.344-16.988f)+
A_s^3(4169.268-901.235f+23.355f^2) \\ \nonumber
C_{F_2,NS}^{(6)}&=&1+11.177A_s+A_s^2(302.399-28.013f)+A_s^3(10069.631
-1816.323f+
42.663f^2) \\ \nonumber
C_{F_2,NS}^{(8)}&=&1+15.530A_s+A_s^2(470.807-37.925f)+A_s^3(17162.372-2787.298f
+61.9118f^2) \\ \nonumber
C_{F_2,NS}^{(10)}&=&1+19.301A_s+A_s^2(639.211-46.861f)+A_s^3(24953.135-
3770.102f+80.5201f^2) \\ \nonumber
C_{F_2,NS}^{(12)}&=&1+22.628A_s+A_s^2(804.585-54.994f)+A_s^3(33171.455-4746.441f
+98.348f^2)
 \\ \nonumber
C_{F_2,NS}^{(14)}&=&1+25.611A_s+A_s^2(965.813-62.465f)+A_s^3(41657.116-
5708.216f+115.392f^2)~~~.
\end{eqnarray}
The NNA versions of the expressions from Eq.~(24) read:
\begin{eqnarray}
C_{F_2,NS}^{(2)}&=&1+0.444A_s+A_s^2(6\times5.333
\tilde{\beta}_0)+
A_s^3(36\times6.030\tilde{\beta}_0^2) \\ \nonumber
C_{F_2,NS}^{(4)}&=&1+6.607A_s+A_s^2(6\times16.988\tilde{\beta}_0)+
A_s^3(36\times23.355\tilde{\beta}_0^2) \\ \nonumber
C_{F_2,NS}^{(6)}&=&1+11.177A_s+A_s^2(6\times28.013\tilde{\beta}_0)+A_s^3(
36\times
42.663\tilde{\beta}_0^2) \\ \nonumber
C_{F_2,NS}^{(8)}&=&1+15.530A_s+A_s^2(6\times37.925\tilde{\beta}_0)+A_s^3(
36\times61.9118\tilde{\beta}_0^2) \\ \nonumber
C_{F_2,NS}^{(10)}&=&1+19.301A_s+A_s^2(6\times46.861\tilde{\beta}_0)+A_s^3(
36\times80.5201\tilde{\beta}_0^2) \\ \nonumber
C_{F_2,NS}^{(12)}&=&1+22.628A_s+A_s^2(6\times54.994\tilde{\beta}_0)+A_s^3(
36\times98.348\tilde{\beta}_0^2)
 \\ \nonumber
C_{F_2,NS}^{(14)}&=&1+25.611A_s+A_s^2(6\times62.465\tilde{\beta}_0)+A_s^3(
36\times115.392\tilde{\beta}_0^2)
\end{eqnarray}
The numerical values of the ratios of the coefficients of Eqs.~(24,25),
namely
$R_{F_2,NNA}^{(2)}(n)=
C_{F_2,NS}^{(2)}(n)_{NNA}/C_{F_2,NS}^{(2)}(n)$ and
$R_{F_2,NNA}^{(3)}(n)=C_{F_2,NS}^{(3)}(n)_{NNA}/C_{F_2,NS}^{(3)}(n)|_{wts}$
are given below. In Table 10 $R_{F_2,NNA}^{(3)}(n)$ is compared
with $R_{F_2}^{(3)}(n)|_{[Pade]}=
C_{F_2}^{(3)}(n)|_{[1/1]}/C_{F_2}^{(3)}(n)|_{wts}$.
Getting support from the related results for $n=10,12,14$ NS moments of $F_2$,
we can make the conclusion that the findings of Ref.~\cite{Mankiewicz:1997gz}
and the new numbers for the moments
of $xF_3$   (see   Tables 7,8) demonstrate
 that at the NNLO and  N$^3$LO the
NNA approximation is working  in the NS channel for $n=1$ and
 $n\geq6$,
which  corresponds to the region of $x$, closer to the
limit $x=1$. In the cases
of low NS  moments the reason for failure  of the NNA approximation
remains unclear. A similar conclusion
can be drawn  from the comparison of the [1/1] Pad\'e estimates for
$C_{F_3}^{(3)}$-terms with the results of the interpolation procedure
(see Table 2 and related discussions after it) and
the $f$-dependence  of the ratios
$R_{F_3,NNA}^{(3)}(n)$ vs $R_{F_3}^{(3)}(n)|_{[Pade]}$
  and $R_{F_2,NNA}^{(3)}(n)$ vs  $R_{F_2}^{(3)}(n)|_{[Pade]}$
 (see Tables 8, 10).

\begin{center}
\begin{tabular}{||r|c|c|c|c||} \hline
$n$& $f=3$ & $f=4$ & $f=5$
 \\ \hline
2& 43.24 &   -18.17    &  -3.27 \\
4& 2.54 &  2.89 & 3.43 \\
6 & 1.74 &  1.83 & 1.98 \\
8&  1.43 &  1.48 & 1.55 \\
10&  1.28 &  1.29 & 1.33 \\
12& 1.16 &  1.18 & 1.19 \\
14& 1.08 &  1.09 & 1.10 \\
\hline
\end{tabular}
\end{center}
{{\bf Table 9.} The $f$-dependence of  the ratios
$R_{F_2,NNA}^{(2)}(n)$. }

\newpage

\begin{center}
\begin{tabular}{||r|c|c|c|c||} \hline
$n$
& $f=3$ & $f=4$ & $f=5$
 \\ \hline
2& 773.92 ({\bf 4.56}) &  -7.75 ({\bf -0.24})    &  -3.41 ({\bf -0.22}) \\
4& 2.54 ({\bf 0.73}) &  3.89 ({\bf 0.87})  & 12.51 ({\bf 1.95})  \\
6 & 1.55 ({\bf 0.85}) &  1.19 ({\bf 0.93}) & 2.74 ({\bf 1.95})\\
8 &  1.18 ({\bf 0.88})&  1.38 ({\bf 0.94})& 1.71 ({\bf 1.07}) \\
10&  1.00 ({\bf 0.90}) &  1.13 ({\bf 0.95})& 1.31 ({\bf 1.05}) \\
12& 0.90 ({\bf 0.91}) &  1.02 ({\bf 0.96}) & 1.09 ({\bf 1.04}) \\
14& 0.82 ({\bf 0.93})  &  0.87 ({\bf 0.97})  & 0.95 ({\bf 1.04}) \\
\hline
\end{tabular}
\end{center}
{{\bf Table 10.} The $f$-dependence of the ratios
$R_{F_2,NNA}^{(3)}(n)$ and $R_{F_2}^{(3)}(n)|_{[Pade]}$.}

 This interesting
similarity might be explained by the studies
of the  large-$\beta_0$ limit of the Pad\'e approximant approach
Ref.~\cite{Brodsky:1997vq} and its
relation to the BLM approach \cite{Brodsky:1983gc}.
Note in these circumstances that
 within the  large-$\beta_0$ limit, the BLM approach
was extended  to all orders in  perturbation theory in Ref.\cite{Ball:1995ni}, while
the possibility of incorporation
of the subleading terms in number of flavours $f$  and constructing the   NNLO
generalization of the BLM approach was first demonstrated in
Ref.~\cite{Grunberg:1992ac} (for further related analysis
see Ref.~\cite{Rathsman:1996jk}).
On the other hand the qualitative sucess of application of
[1/1] Pad\'e approximation for $n\geq 6$ might be considered as
the additional argument in favour of possibility of its
application for estimating N$^3$LO contribution $r(n)$ to the
expanded anomalous dimension term Eq.~(10).

\subsection{The determination of $\alpha_s(M_Z)$ values  and
\\ their  scale-dependence uncertainties}

As  is known from the work of Refs.~\cite {vanNeerven:2000ca,Neerven:2001pe}
it is rather instructive to consider  the sensitivity of the
results of the perturbative   QCD analysis   to the
variation of renormalization and factorization scales.
We will  study the question of factorization-renormalization
scale dependence within the class of $\overline{\rm MS}$-like schemes.
This means that we will change
 only the scales without varying the scheme-dependent
coefficients of anomalous dimensions and
$\beta$-function.

The  arbitrary factorization scale  enters in the following
equation:
\begin{equation}
A_s( Q^2/\mu^2_{\overline{MS}})= A_s(Q^2/\mu_F^2)\bigg[1+k_1 A_s(Q^2/\mu_F^2)+
k_2 A_s^2(Q^2/\mu_F^2)+k_3A_s^3(Q^2/\mu_F^2)  \bigg]
\end{equation}
where $\mu_F^2$  is the factorization scale
and
\begin{eqnarray}
k_1&=&\beta_0 ln(\frac{\mu^2_{\overline{MS}}}{\mu_F^2}) \\ \nonumber
k_2&=&k_1^2+\frac{\beta_1}{\beta_0}k_1 \\ \nonumber
k_3&=&k_1^3+\frac{5\beta_1}{2\beta_0}k_1^2+\frac{\beta_2}{\beta_0}k_1
\end{eqnarray}
Let us choose the factorization scale as $\mu_F^2=\mu_{\overline{MS}}^2 k_F$.

Then we have:
\begin{equation}
k_1=-\beta_0 ln(k_F)
\end{equation}

In this case after application of the renormalization group
equation and substitution of Eq.~(26) into Eqs.~(9,10) of Sec.2
we get
\begin{equation}
exp\bigg[-\int^{A_s(Q^2/ k_F)}\frac{\gamma_{NS}^{(n)}(x)}{\beta(x)}dx\bigg]
=(A_s(Q^2/k_F))^{a}\times \overline{AD}(n,A_s(Q^2/ k_F))
\end{equation}
where $a=\gamma_{NS}^{(0)}/2\beta_0$ and
\begin{eqnarray}
\overline{AD}(n,A_s(Q^2/ k_F))&=&1+\bigg[p(n)+a k_1\bigg]A_s(Q^2/ k_F) \\ \nonumber
&+&\bigg[q(n)+p(n)k_1(a+1)+\frac{\beta_1}{\beta_0}k_1
a+\frac{a(a+1)}{2}k_1^2\bigg]A_s^2(Q^2/ k_F) \\ \nonumber
&+&\bigg[r(n)+q(n)k_1(a+2)+p(n)\left(\frac{\beta_1}{\beta_0}k_1(a+1)
+\frac{(a+1)(a+2)}{2}k_1^2\right) \\ \nonumber
&+& \frac{\beta_2}{\beta_0}k_1a+\frac{\beta_1}{\beta_0}k_1^2a(\frac{3}{2}+a)+
\frac{a(a+1)(a+2)}{6}k_1^3\bigg]A_s^3(Q^2/ k_F)
\end{eqnarray}

Now let us consider  the factorization and renormalization scale dependence,
fixing  $k_R=k_F=k$.
In this case we should also  modify the coefficient function in Eq.(6) as
\begin{eqnarray}
C_{F_3}^{(n)}&=&1+C^{(1)}_{F_3}(n)A_s(Q^2/ k)+\bigg[C^{(2)}_{F_3}(n)-
C^{(1)}_{F_3}(n)\beta_0 ln(k)
\bigg]A_s^2(Q^2/ k) \\ \nonumber
&+& \bigg[C^{(3)}_{F_3}(n)+C^{(1)}_{F_3}(n)
\left(\beta_0^2ln^2(k)-\beta_1ln(k)\right)
-2C^{(2)}_{F_3}\beta_0ln(k)\bigg]A_s^3(Q^2/ k)~~~~~.
\end{eqnarray}

The commonly accepted practice is to   vary $k$
in the interval $1/4\leq k\leq 4$ (see, e.g., Ref.\cite{vanNeerven:2000ca}).
We repeated our fits both without
and with the IRR model of the twist-4 terms
in the cases of $k=1/4$ and $k=4$. As in the fits
described above, in order to achieve the minimum in $\chi^2$
we supplemented the interpolation
procedure of the NNLO approximation for
$\gamma_{F_3}^{(n)}(A_s)$ by  fine-tuning of even terms
$\gamma_{F_3}^{(2)}(6)$,  $\gamma_{F_3}^{(2)}(8)$ and  $\gamma_{F_3}^{(2)}(10)$
and got their values, comparable within small error-bars
with the numbers given in Sec. 4.
The same procedure was used in the process of the  N$^3$LO fits.
In fact they  have more theoretical uncertainties than
the NNLO ones. Indeed, in this case we applied the interpolation procedure
to determine not only the NNLO coefficients of anomalous dimensions
of even moments of $xF_3$ SF, but the
related N$^3$LO terms of the coefficient functions as well (see Table 2).
The N$^3$LO   corrections $r(n)$ to $AD$-function  in Eq.~(30)
were modelled using the  [1/1] Pad\'e approximant procedure.
Note that for  even values of $n$  the numerical expresssions for
$q(n)$, which enter into the Pad\'e approximants,
are determined in part by the NNLO coefficients
of anomalous dimensions of even moments of $xF_3$.
In the process of N$^3$LO fits in the case of $k=4$ the ambiguities of the
applications of the [1/1] Pad\'e approximation
procedure  reflect themselves  in the necessity of supplementing the
interpolation procedure by fine-tuning
of the coefficient $\gamma_{F_3}^{(2)}(12)$ in addition
to the $n=6,8,10$ NNLO anomalous dimension terms.
Only after this additional step
were we able to achieve a
reasonable value of $\chi^2$ in this case also.

The  consequences   of the  study of factorization/renormalization
scale dependence at the  NLO, NNLO and N$^3$LO
in the case of the initial scale $Q_0^2=20$ GeV$^2$
are presented in Table 11.

\begin{center}
\begin{tabular}{||r|r|r|r|r|c|c|}
\hline
   Order & $N_{max}$ & k & $\Lambda_{\overline{\rm MS}}^{(4)}$ &  $\Delta_k$      & $A'_2~(HT)$  &  $\chi^2$/points       \\
\hline
   NLO   &9 &   1   &   345$\pm$38 & ---     &       ---    &    85.1/86       \\
         &9 &   1    &  376$\pm$39 & ---        &    - 0.121$\pm$0.052  &
79.5/86 \\

         &9 &   1/4    & 482$\pm$57 &  137      &       ---    &
90.0/86       \\
         &9 &   1/4    & 579$\pm$62 & 203         &    - 0.184$\pm$0.054
&    78.8/86       \\
         &9 &   4  &  270$\pm$25 &  -75    &       ---       &
84.7/86       \\
         &9 &   4  &  271$\pm$24 & -105       &    - 0.032$\pm$0.051   &
84.4/86        \\
\hline\hline
NNLO & 9 & 1 & 331$\pm$37  & --- & --- &  75.8/86 \\
     & 9 & 1 & 332$\pm$36  & --- & -0.015$\pm$0.051 & 75.7/86 \\
     & 9&   1/4 &379$\pm$45 &      47        &       ---       &   78.7 /86       \\

      & 9&   1/4    &  399$\pm$46  & 66        &    - 0.084$\pm$0.052
&    76.1/86        \\
         & 9 & 4  & 297$\pm$27 &   -35       &       ---       &    79.4/86       \\
         & 9  & 4  & 318$\pm$30 &  -15         &    + 0.117$\pm$0.052
&    74.9/86       \\
\hline\hline
   N$^3$LO & 9 & 1 & 327$\pm$34 & --- & --- &  76.4/86 \\
           & 9 & 1 & 329$\pm$34 & --- & +0.033$\pm$0.052 & 76.0/86 \\
 & 9&   1/4    & 355$\pm$39 &    28       &       ---       &   75.9 /86       \\
         & 9&   1/4    & 357$\pm$39 & 28         &    - 0.026$\pm$0.051  &
75.9/86        \\
         & 9 & 4  & 312$\pm$24 &  -15       &       ---       &    74.8/86       \\
         & 9  & 4  & 318$\pm$24 & -11         &    + 0.058$\pm$0.052
&    84.5/86       \\
\hline
\end{tabular}
\end{center}
{{\bf Table 11.}
The outcomes of NLO, NNLO and N$^3$LO fits to  CCFR'97 $xF_3$ data for
$Q^2\geq 5~GeV^2$ with different values
of factorization/renormalization scales.
The difference in the values of
$\Lambda_{\overline{\rm MS}}^{(4)}$
is determined by $\Delta_k (MeV)=\Lambda^{(4)}_{\overline{\rm MS}}(k)
-\Lambda^{(4)}_{\overline{\rm MS}}(k=1)$. The value of the
IRR model coefficient is given in GeV$^2$.
The initial scale is fixed at $Q_0^2=20~$GeV$^2$.}

It should be  mentioned that despite  the approximate nature
of the Pad\'e resummation procedure used for the estimation
of the N$^3$LO contribution $r(n)$ to the  anomalous dimension function
$AD$, in the case of  application of the  fine-tuning procedure at
 $k=1$ and $k=1/4$ we  get
for  $\gamma_{F_3}^{(2)}(n)$ with $n=6,8,10,12$ the
numerical expressions
which agree with the interpolated numbers of Table 1 in the 4th
significant digit.
For the
 fine-tuned fits  with $k=4$ and  high-twist terms included
the price for small values of $\chi^2$ is paid by more
approximate determinations of
$\gamma_{F_3}^{(2)}(10)$ and
$\gamma_{F_3}^{(2)}(12)$ which differ from the
ones presented in Sec. 4  in the 3rd  significant digit.
For example, for $n=12$ the following
number was obtained: $\gamma_{F_3}^{(2)}(12)\approx 1694.4907\pm 3.1194$.
It should be stressed, however, that this difference leads to
negligibly small effects in the overall contributions
to Pad\'e estimated values of $r(n)$. Moreover, it is rather
pleasant that  the  approximate character
of the fixation  of this part of the  theoretical input of  our
new N$^3$LO fits
does not  drastically spoil  reasonable (from our point of view)
estimates for
$\gamma_{F_3}^{(2)}(n)$ at  $n=6,8,10,12$.

We make  now several
conclusions which come from the results presented in Table 11.
\begin{itemize}
\item At NLO, NNLO and N$^3$LO  and $k=1/4$ the values of
$\Lambda_{\overline{\rm MS}}^{(4)}$ and thus $\alpha_s$ are
larger than in the case of $k=1$, while for $k=4$  smaller
numbers are obtained.
\item The NLO and NNLO results of Table 11 are in
satisfactory agreement with the similar ones from Table 6 of Ref.~\cite{KPS2},
provided one takes into account the difference in the definitions
of the parameter $k$ (in Ref.~\cite{KPS2} the case $k=4$ ($k=1/4$)
corresponds to the choice  $k=1/4$ ($k=4$) in Table 11).
\item To our point of view the results  of the NNLO fits
with $k=1/4$ ($k=4$) both without and with twist-4 terms
simulate in part the results of the NLO (N$^3$LO) fits
with  $k=1$.
\item The increase of order of perturbative theory approximations
leads to minimization of the scale-dependence uncertainty
which manifests itself through the decrease of
the values of $\Delta_k$ deviations.
\item In the case of NLO fits with HT terms the
value of $|\Delta_k|$ is larger than in the case of
switching of power-supressed terms. However, this difference
is minimized at the NNLO and the the N$^3$LO especially.
We think that this property is reflecting the correlation with
the effective minimization of the fitted value of the HT parameter
$A_2^{'}$, which   becomes comparable with zero in the NNLO and N$^3$LO
fits with $k=1$.
\item We checked that for $k=1$ and $k=1/4$ the results are not sensitive
to the changes  of the initial scale from $Q_0^2=20$ GeV$^2$ to
$Q_0^2=5$ GeV$^2$.
\item However, when $k=4$, this pleasant feature
is violated  in the results of the   NNLO
and N$^3$LO fits especially. Indeed, these fits  are
accompanied by the increase of
$\chi^2$ up to over the  100/86 level. This fact  can be  related to
pushing  the value of $Q_0^2$ out of  the considered
kinematical region $Q^2\geq 5$
GeV$^2$  (at $Q_0^2=5$ GeV$^2$ and $k=4$ the region
of $1.25$ GeV$^2\leq Q^2\leq 5$ GeV$^2$
should be also taken into account; however
 the NNLO and N$^3$LO corrections
to the coefficient functions are rather large in this region).
\item It should be stressed that the NLO and NNLO results of Table 11
with $k=4$ are closer to the ones obtained in the recent
work of Ref.~\cite{Santiago:2001mh}. We will comment on the
possible consequences of this observation in Sec.6.2
below.
\end{itemize}

Let us now turn to
determination of  the values of $\alpha_s(M_Z)$ from the results
of Table 11.  We   transform $\Lambda_{\overline{\rm MS}}^{(4)}$
into $\Lambda_{\overline{\rm MS}}^{(5)}$ using the NLO, NNLO and  N$^3$LO
variants of equation of Ref.~\cite{Chetyrkin:1997sg}:
 \begin{eqnarray}
\beta_0^{f+1}ln\frac{\Lambda
_{\overline{MS}}^{(f+1)~2}}{\Lambda_{\overline{MS}}^{(f)~2}}=
(\beta_0^{f+1}-\beta_0^f)L_h \\ \nonumber
+\delta_{NLO}+\delta_{NNLO}+
\delta_{N^3LO}
\end{eqnarray}
\begin{equation}
\delta_{NLO}=\bigg(\frac{\beta_1^{f+1}}{\beta_0^{f+1}}-\frac{\beta_1^f}
{\beta_0^f}\bigg)ln L_h-\frac{\beta_1^{f+1}}{\beta_0^{f+1}}ln
\frac{\beta_0^{f+1}}{\beta_0^f}
\end{equation}
\begin{eqnarray}
\delta_{NNLO}=\frac{1}{\beta_0^f L_h}\bigg[\frac{\beta_1^f}{\beta_0^f}
\bigg(\frac{\beta_1^{f+1}}{\beta_0^{f+1}}
-\frac{\beta_1^f}{\beta_0^f}\bigg)ln L_h
\\ \nonumber
+\bigg(\frac{\beta_1^{f+1}}{\beta_0^{f+1}}\bigg)^2
-\bigg(\frac{\beta_1^{f}}{\beta_0^{f}}\bigg)^2
-\frac{\beta_2^{f+1}}{\beta_0^{f+1}}
+\frac{\beta_2^{f}}{\beta_0^{f}}-C_2\bigg]
\end{eqnarray}
\begin{eqnarray}
\delta_{N^3LO}=\frac{1}{(\beta_0^f L_h)^2}\Bigg[
-\frac{1}{2}\bigg(\frac{\beta_1^f}{\beta_0^f}\bigg)^2
\bigg(\frac{\beta_1^{f+1}}{\beta_0^{f+1}}-\frac{\beta_1^f}{\beta_0^f}\bigg)
ln^2 L_h
\\ \nonumber
+\frac{\beta_1^f}{\beta_0^f}\bigg[-\frac{\beta_1^{f+1}}{\beta_0^{f+1}}
\bigg(\frac{\beta_1^{f+1}}{\beta_0^{f+1}}-\frac{\beta_1^f}{\beta_0^f}\bigg)
+\frac{\beta_2^{f+1}}{\beta_0^{f+1}}
-\frac{\beta_2^f}{\beta_0^f}+C_2\bigg]ln L_h
\\ \nonumber
+\frac{1}{2}\bigg(-\bigg(\frac{\beta_1^{f+1}}{\beta_0^{f+1}}\bigg)^3
-\bigg(\frac{\beta_1^f}{\beta_0^f}\bigg)^3-\frac{\beta_3^{f+1}}{\beta_0^{f+1}}
+\frac{\beta_3^f}{\beta_0^f}\bigg)
\\ \nonumber
+\frac{\beta_1^{f+1}}{\beta_0^{f+1}}\bigg(\bigg(
\frac{\beta_1^f}{\beta_0^f}\bigg)^2+
\frac{\beta_2^{f+1}}{\beta_0^{f+1}}-\frac{\beta_2^f}{\beta_0^f}+C_2\bigg)
-C_3\Bigg]
\end{eqnarray}
where
$\beta_i^f$ ($\beta_i^{f+1}$) are the  coefficients of the
$\beta$-function with $f$ ($f+1$)
numbers of active flavours,
$L_h=ln(M_{f+1}^2/\Lambda_{\overline{MS}}^{(f)~2})$ and $M_{f+1}$ is the
threshold of the production of a quark of the $(f+1)$ flavour and
$C_3=-(80507/27648)\zeta(3)-(2/3)\zeta(2)
((1/3) ln2+1)-58933/124416 +(f/9)[\zeta(2)+2479/3456]$.

These formulae contain the
NNLO correction to the matching
condition with coefficient $C_2=-7/24$, previously derived in Ref.~\cite{Bernreuther:1982sg} and
correctly calculated in
Ref.~\cite{Larin:1995va}.
In our massless analysis we will take $f=4$ and $m_b\approx 4.8~GeV$
and  vary the threshold of the production of the fifth
flavour from $M_5^2\approx m_b^2$ to $M_5^2\approx(6m_b)^2$ in accordance
with the proposal of Ref.~\cite{Blumlein:1999sh}. The latter choice
is based on the calculations of the LO and NLO
massive corrections to the Gross-Llewellyn Smith sum rule.
The final values of $\alpha_s(M_Z)$ will be fixed  at the middle
of the interval, limited by the
choices of threshold
matching point at $M_5^2\approx m_b^2$ and $M_5^2\approx(6m_b)^2$.
The appearing  theoretical ambiguities
reflect the uncertainties due to the
manifestation  of the massive-dependent contributions
to the moments of $xF_3$ in the massless fits.
Another procedure of fixing the massive-dependent
ambiguities in $\alpha_s(M_Z)$, which result from the fits
to $xF_3$ data, was proposed in Ref.~\cite{Shirkov:1997h}.
It is based on the  application  of the
massive-dependence of the LO contribution to $\beta$-function in
the MOM scheme, previously studied in the works of Refs.~\cite{Shirkov:1992pc,
Shirkov:1994td}. This procedure gives the
estimates of the influence of mass-dependence on the value
of $\alpha_s(M_Z)$, extracted from  CCFR'97 $xF_3$ data, which
are comparable to ours.
To be more complete at this point, we also mention several other works,
which are dealing with different
prescriptions for estimating threshold uncertainties (see  Refs.
\cite{Peterman:1992uj,Brodsky:1998mf} and the work of
Ref.~\cite{Jegerlehner:1999zg} especially, where massive-dependence
of the MOM-scheme
coupling constant was evaluated at the 2-loop level). It was shown in
Ref.~\cite{Blumlein:1999sh} that the application of the $\overline{\rm MS}$-scheme matching condition with the matching point $M_5^2\approx (6m_b)^2$
does not contradict the  application of
the massive dependent approach of Ref.~\cite{Jegerlehner:1999zg}.
Therefore, we can conclude that our estimates of massive-dependent
uncertainties in $\alpha_s(M_Z)$ can be substantiated by this comparisons
of the results of
Refs.~\cite{Blumlein:1999sh,Jegerlehner:1999zg}.

Taking into account the numbers given in  Table 11, which were
obtained with the  twist-4 contribution fixed using  the
IRR model  of Ref.~\cite{DW}, and the
theoretical expressions of Eqs.~(14)-(16) and Eqs.~(32)-(35), supplemented
with the  estimates of the uncertainties due to different
possibilities of the choice of matching point and experimental systematic
errors, which come from separate consideration of this type
of experimental uncertainties of CCFR'97 data,   we arrive at the
following values of $\alpha_s(M_Z)$, extracted from
the fits to CCFR'97 data for $xF_3$ performed in this work:
\begin{eqnarray}
NLO~HT~of~ Ref.\cite{DW}~\alpha_s(M_Z)&=&0.120 \pm 0.0022(stat)
\pm 0.005(syst) \\ \nonumber
&&\pm 0.002(thresh.)^{+0.010}_{-0.006}(scale)
\end{eqnarray}
\begin{eqnarray}
NNLO~HT~of~ Ref.\cite{DW}~\alpha_s(M_Z)&=&0.119\pm 0.002(stat)
\pm 0.005(syst) \\ \nonumber
&&\pm 0.002(thresh.)^{+0.004}_{-0.002}(scale)
\end{eqnarray}

Minor differences with the similar results
of Ref.~\cite{KPS2} are explained by the  incorporation of
more significant digits in the process of calculations
and by  more careful study of the scale-dependence uncertainties.
These values presented in Eqs.~(36,37)
should be compared with the one given by
our new N$^3$LO approximate
fit:
\begin{eqnarray}
N^3LO~HT~of~ Ref.\cite{DW}~\alpha_s(M_Z)&=&0.119\pm 0.002(stat)
\pm 0.005(syst) \\ \nonumber
&&\pm 0.002(thresh.)^{+0.002}_{-0.001}(scale)~~~.
\end{eqnarray}

Note again, that the experimental systematic uncertainties are extracted
from the CCFR'97 data and were not taken into account in the process
of our concrete studies for the reasons discussed above.
As to the theoretical uncertainties of the  $\alpha_s(M_Z)$-value,
the incorporation of the
high-order corrections
to the fits   leaves
threshold ambiguities  at  the same level, but  decreases
scale-dependence uncertainties drastically
\footnote{However,  like
in other studies of the CCFR'97 data of
Refs.~\cite{KKPS2}-\cite{KPS2},~\cite{CCFR2},~\cite{Santiago:2001mh}-
\cite{AK2},~\cite{Shirkov:1997h}, we are neglecting  theoretical
uncertainties, which arise
from the   separation  from
the CCFR'97 data heavy nuclei corrections.
Definite theoretical considerations
of this problem   \cite{Kulagin:1998vv,Kulagin:1998wc} were incorporated
in the fits in Ref.~\cite{Kulagin:2000yw} and
indicate  the decrease of $\alpha_s(M_Z)$
at the NNLO level to the amount   about $2\times 10^{-3}$.}.

To study  the influence of  the twist-4 contributions
of Eq.~(21) to the values of $\alpha_s(M_Z)$ we  also extracted
from Table 11 the corresponding results, obtained from the twist-4
independent fits to CCFR'97 data:
\begin{eqnarray}
NLO~\alpha_s(M_Z)&=&0.118 \pm 0.002(stat)
\pm 0.005(syst) \\ \nonumber
&&\pm 0.002(thresh.)^{+0.007}_{-0.005}(scale)
\end{eqnarray}
\begin{eqnarray}
NNLO~\alpha_s(M_Z)&=&0.119\pm 0.002(stat)
\pm 0.005(syst) \\ \nonumber
&&\pm 0.002(thresh.)^{+0.003}_{-0.002}(scale)~~~.
\end{eqnarray}
The approximate N$^3$LO twist-independent fits give us  the following
numbers:
\begin{eqnarray}
N^3LO~\alpha_s(M_Z)&=&0.119\pm 0.002(stat)
\pm 0.005(syst) \\ \nonumber
&&\pm 0.002(thresh.)^{+0.002}_{-0.001}(scale)
\end{eqnarray}
Let us make several conclusions, which follow from the
comparison of the results
of Eqs.~(36)-(38) and Eqs.~(39)-(40).
\begin{itemize}
\item In the case when HT-corrections are included, the general
tendency $(\alpha_s(M_Z))_{NLO}\geq(\alpha_s(M_Z))_{NNLO}\geq
(\alpha_s(M_Z))_{N^3LO}$ for the central values of
the outcomes of the fits takes place.
\item The scale dependence of the NLO and NNLO results
with HT-corrections included are larger than in the case
of $\alpha_s(M_Z)$-values, obtained without HT-terms.
\item The scale-dependence of the results of the approximate N$^3$LO fits
both without and with HT-corrections is almost the same. This
feature is related to effective minimization of the
contributions of HT-terms at the N$^3$LO.
\item Starting from the NNLO the systematical experimental uncertainties
dominate  theoretical ambiguities as estimated by us.
\item The uncertainties of the matching conditions
dominate  scale-dependence ambiguities at the N$^3$LO only.
This might mean
that the approximation of massless quarks
works reasonably well
in the analysis of CCFR'97   $xF_3$  data up to NNLO.
\end{itemize}

It is worth making several comments on the comparison
of our results for $\alpha_s(M_Z)$, extracted from
the CCFR'97 $xF_3$ data,  with those available in the literature.
Within existing theoretical
error bars, which reflect  in part the  special features of the
procedures
of taking into account threshold effects,
our twist-independent NLO result is  in agreement
with  the value  $\alpha_s(M_Z)_{NLO}\approx 0.122 \pm 0.004$, given
by   the independent  NLO fits to CCFR'97 $xF_3$ data
using the Jacobi polynomial
method \cite{Shirkov:1997h}. It is also in agreement with the
one presented by the
CCFR collaboration, namely  $\alpha_s(M_Z)_{NLO}=0.119 \pm
0.002 (exp) \pm 0.004 (theory)$ \cite{CCFR2}, which was obtained with the help of the
DGLAP approach and reproduced in Ref.~\cite{Alekhin:1999df} without
treating carefully theoretical uncertainties.
Within experimental and theoretical error bars
our results also do not contradict  the recent application
of the Bernstein polynomial method \cite{Yndurain:1978wz}
for the extraction of $\alpha_s(M_Z)$ from CCFR'97 $xF_3$ data
at the NLO and NNLO~\cite{Santiago:2001mh}. Indeed, at the NLO
it gives $\alpha_s(M_Z)_{NLO}=0.116\pm 0.004$, while
at the NNLO they got  $\alpha_s(M_Z)=0.1153 \pm 0.0041 (exp)
\pm 0.0061 (theor)$ \cite{Santiago:2001mh}.
It is worth mentioning  here that despite the qualitative agreement with
our results, the
central values of
$\alpha_s(M_Z)$ obtained in Ref.~\cite{Santiago:2001mh}  are
lower than the central values of all
existing NLO and NNLO determinations of $\alpha_s(M_Z)$ from CCFR'97
$xF_3$ data. In Section 6 we will present  more detailed comparison of
the results
of Ref.~\cite{Santiago:2001mh} with the ones
obtained in our work and will
propose a  possible explanation of the {\bf origin} of these deviations.

Although we do not include in our fits
the simultaneous analysis of the statistical and systematic experimental
uncertainties, we think that our analysis  has some  advantages over
other fits to CCFR'97 $xF_3$ data. Indeed, the results  of
Refs.~\cite{Shirkov:1997h,Santiago:2001mh} are free
from considerations of the scale-dependence ambiguities studied  in our work,
while the DGLAP fits of the same data, performed in
Refs.~\cite{CCFR2,Alekhin:1999df,AK2}, do not take  into consideration
the contributions of the NNLO and N$^3$LO perturbative QCD corrections,
which we were able to treat in the way  described above.

To our point of view, another advantage of our work
is that using more {\bf rigorous} information  than
previously (see Ref.~\cite{KPS2})
on the effects of the higher-order perturbative QCD contributions
to the characteristics of $xF_3$ SF, we continued the studies of the
influence of
twist-4 corrections to the extraction of $\alpha_s(M_Z)$ from CCFR'97
$xF_3$ data. However, the  IRR approach used by us  is not the only
way of modelling twist-4 effects. In the next Section
we are considering the case when the  twist-4 contribution will be
approximated
in a  less  model-dependent way.

\subsection{Determination of the $x$-shape of the  twist-4 corrections}

We now turn to a pure phenomenological extraction of the twist-4 contribution
$h(x)$ to Eq.~(18). In order to get its $x$-dependence we will model $h(x)$
by free parameters $h_i=h(x_i)$, where $x_i$ are the points in experimental
data binning. The results are presented in Table 12.

\begin{center}
\begin{tabular}{||c|c|c|c|c||} \hline \hline
                                        &      LO                  &
NLO                  &    NNLO               &           N$^3$LO
\\ \hline
 $\chi^2/{nep }$                        &     69.1/86              &
68.5/86              &   65.6/86             &    66.1/86                \\
   A                                    &     4.32  $\pm$   1.34   &
4.08  $\pm$   1.13  &    5.06  $\pm$  0.46  &     5.55  $\pm$   1.27    \\
   b                                    &    0.629  $\pm$  0.096   &
0.616  $\pm$  0.084  &   0.682  $\pm$  0.030 &    0.711  $\pm$  0.079     \\
   c                                    &     4.28  $\pm$  0.14    &
4.16  $\pm$  0.15   &    3.88  $\pm$  0.20  &     3.73  $\pm$  0.34      \\
$\gamma$                                &     1.91  $\pm$   1.20   &
1.87  $\pm$   1.09  &   0.73   $\pm$  0.30  &    0.25   $\pm$  0.81     \\
$\Lambda_{\overline{MS}}^{(4)}$ $[MeV]$ &     327   $\pm$   149    &
395   $\pm$   151   &    391   $\pm$   159  &     370   $\pm$   131     \\
$\alpha_s(M_Z)$ & 0.140$^{+0.010}_{-0.010}$ & 0.123$^{+0.007}_{-0.007}$
& 0.124$^{+0.008}_{-0.010}$ & 0.123$^{+0.007}_{-0.008}$ \\
\hline
   $x_i$                     &\multicolumn{4}{c||}{  $h(x_i)~[GeV^2]$ }
\\  \hline
0.0125                                  &    0.016  $\pm$  0.295   &
0.056  $\pm$  0.281  &   0.054  $\pm$  0.274 &    0.072  $\pm$  0.271    \\
0.025                                   &   -0.055  $\pm$  0.241   &
0.038  $\pm$  0.235  &   0.239  $\pm$  0.203 &    0.284  $\pm$  0.228    \\
0.050                                   &   -0.079  $\pm$  0.161   &
0.113  $\pm$  0.206  &   0.425  $\pm$  0.275 &    0.472  $\pm$  0.306    \\
0.090                                   &   -0.231  $\pm$  0.112   &
0.050  $\pm$  0.193  &   0.146  $\pm$  0.283 &    0.158  $\pm$  0.277    \\
0.140                                   &   -0.369  $\pm$  0.108   &
-0.078  $\pm$  0.132  &  -0.012  $\pm$  0.211 &    0.013  $\pm$  0.223    \\
0.225                                   &   -0.492  $\pm$  0.223   &
-0.281  $\pm$  0.160  &  -0.038  $\pm$  0.123 &    0.062  $\pm$  0.163    \\
0.350                                   &   -0.344  $\pm$  0.371   &
-0.347  $\pm$  0.367  &  -0.207  $\pm$  0.303 &   -0.064  $\pm$  0.227    \\
0.550                                   &    0.129  $\pm$  0.304   &
-0.026  $\pm$  0.335  &  -0.172  $\pm$  0.385 &   -0.140  $\pm$  0.325    \\
0.650                                   &    0.398  $\pm$  0.195   &
0.275  $\pm$  0.219  &   0.144  $\pm$  0.282 &    0.131  $\pm$  0.255    \\
 \hline
   n        &&           &  $\gamma_{F_3}^{(2)}(n)$ &
$\gamma_{F_3}^{(2)}(n)$   \\  \hline
   6                                    &                          &
&   1247.9 $\pm$ 0.7 &    1247.7 $\pm$ 0.6    \\
   8                                    &                          &
  &   1419.9 $\pm$ 0.8 &    1419.9 $\pm$ 0.8    \\
  10                                    &                          &
 &   1561.9 $\pm$ 2.2 &    1561.6 $\pm$ 2.2    \\
  12                                    &                          &
 &   1676.7 $\pm$ 5.3 &    1676.8 $\pm$ 4.9    \\
\hline  \hline
\end{tabular}
\end{center}
{{\bf Table 12.} The values for  of the the  parameters
$h(x_i)$, A, b, c, $\gamma$ and
$\Lambda_{\overline{\rm MS}}^{(4)}$
with corresponding statistical errors. They are obtained
from the fits with $N_{max}=9$ and
$Q_0^2=20~{\rm GeV}^2$.}


\begin{figure}[p]
\centerline{ \epsfxsize=5.2in\epsfbox{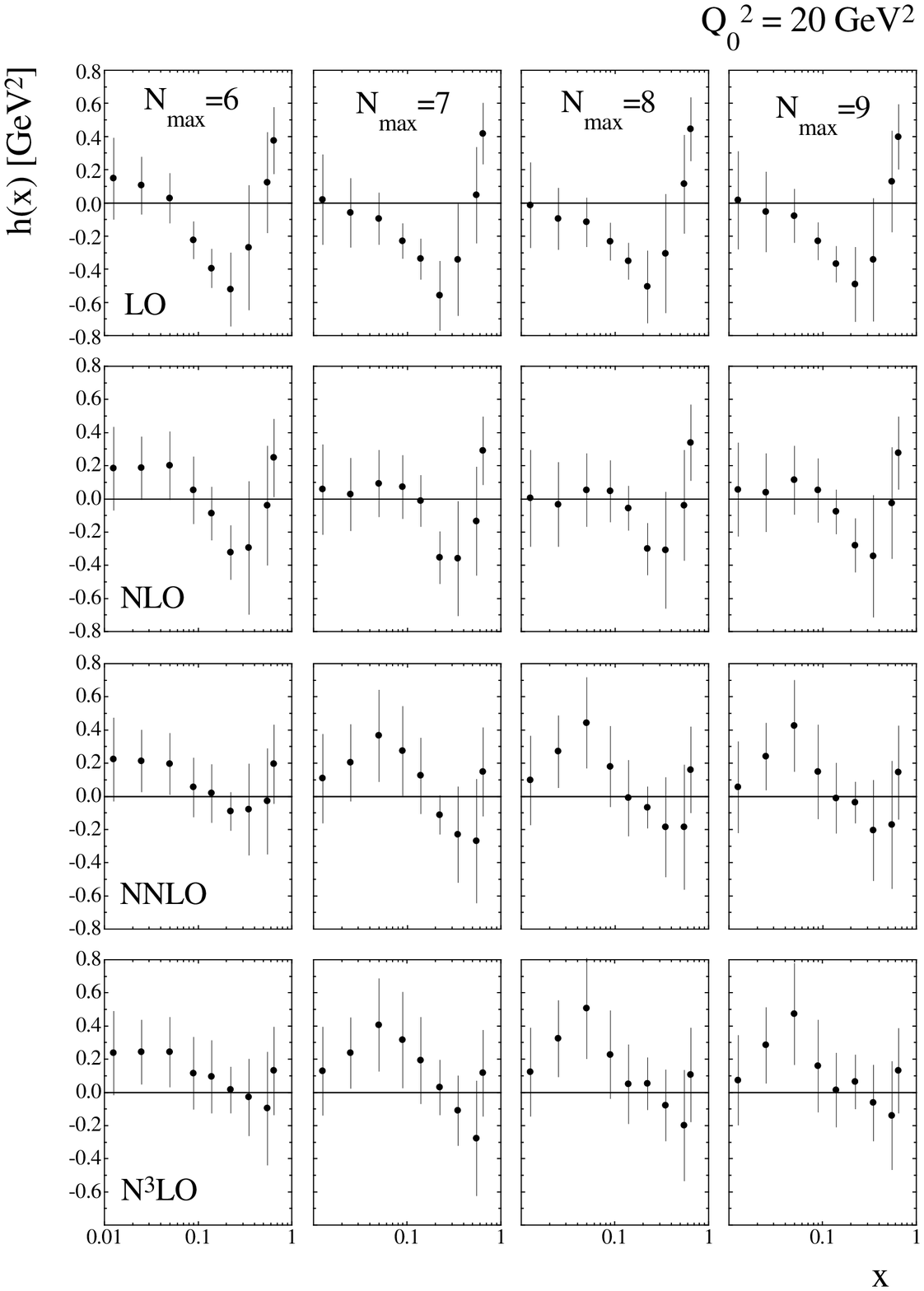} }
 \caption{
The $x$-shapes of $h(x)$ extracted  from the fits to $xF_3$
CCFR'97 data in different orders of perturbative theory. The
initial scale is chosen at $Q_0^2=20~GeV^2$. The cases of
different $N_{max}$ are considered. a) $N_{max}$=6; b) $N_{max}$=7;
c) $N_{max}$= 8; d) $N_{max}$=9}.
\end{figure}

The $x$-shapes of $h(x)$, obtained at LO, NLO, NNLO and approximate
N$^3$LO,  are  depicted at Fig.1, where we also
illustrate the similar behaviour of $h(x)$, obtained in the cases of
$N_{max}=6,7,8$, which correspond to smaller number of Mellin moments,
used in the perturbative part of the Jacobi polynomial reconstruction formula
of  Eq.~(18).

It should be noted that
to minimize correlations between the values of $h_i$,
the parameters of the model for $xF_3(x,Q_0^2)=
A(Q_0^2)x^{b(Q_0^2)}(1-x)^{c(Q_0^2)}(1+\gamma(Q_0^2)x)$ and the QCD scale
$\Lambda_{\overline{\rm MS}}^{(4)}$, we
choose 9 twist-4 parameters $h_i$ only, contrary to 16
ones considered  in the process of our previous analysis of
Refs.~\cite{KKPS2,KPS2}. The results of the fits are presented in
Table 12.
The approximate N$^3$LO fits are based
on the  application of available N$^3$LO corrections to the coefficient
functions of odd  Mellin moments for $xF_3$ \cite{RV}, supplemented
with the smooth interpolation procedure (see Table 2) and using the
[1/1] Pad\'e model of the N$^3$LO contributions $r(n)$ to the expanded anomalous dimension term of Eq.~(10) (see Table 3). In the process of
NNLO and approximate N$^3$LO fits  we faced a problem
identical to the one revealed while fitting $xF_3$ CCFR'97 data without twist-4 terms (see Sec.4) and with twist-4 contributions, modelled by means of the
IRR
approach (see Eqs.~(20,21)). Indeed, to get the stable value of $\chi^2$
at $N_{max}=9$ it was necessary to apply the  fine-tuning procedure for the
NNLO corrections to $\gamma_{F_3}^{(n)}$ at $n=6,8,10,12$.
The obtained values of these parameters are
presented in Table 12 also. Within error-bars they are in agreement
with the numbers fixed by the  smooth interpolation procedure
(see Table 1).

Several comments are  now in order.
\begin{enumerate}
\item The obtained values of $\chi^2$, given in Table 12, are
considerably smaller than the ones obtained in the process
of the fits without twist-4 contributions (see Table 4) and
with $1/Q^2$-terms modelled through the IRR approach (see Table 6).
This is the welcome feature of the analysis of DIS data, which
is based on the model-independent parametrization of the twist-4
terms.
\item The parameters A, b, c, $\gamma$ and
$\Lambda_{\overline{\rm MS}}^{(4)}$ given in Table 12 are in agreement with
their values,
obtained in Ref.~\cite{KPS2} with the help of the fits,
which were made in the case of  $N_{max}=6$  and  16 HT parameters $h_i$.
\item Due to the effect of correlations of $h_i$ and
$\Lambda_{\overline{\rm MS}}^{(4)}$ the values of the
QCD scale parameter have
rather large statistical error-bars.
\item At the LO and NLO the $x$-shape for  $h(x)$ is
rather stable to the increase
of $N_{max}$ from $N_{max}=6$ to $N_{max}=9$ and therefore, to  the
incorporation of the
additional Mellin moments in the procedure of reconstruction of $xF_3$
via  the Jacobi polynomial technique.
\item The $x$-shape of $h(x)$, obtained at the LO and NLO, is in agreement with the prediction of the IRR model of Ref.~\cite{DW}.
\item At the NNLO we observe the sinusoidal-type oscillations of $h(x)$, which
are becomimg more vivid in the cases of $N_{max}=8$ and $N_{max}=9$.
\item This feature seems to be in agreement with the qualitative expectations
which result from an educated guess about the possible modifications
of the prediction of the IRR model at the NNLO~\cite{Webber}.
\item At  the N$^3$LO, for all considered $N_{max}$, the shapes of $h(x)$ are
similar to the ones obtained in the process of new NNLO fits.
\item It is worth  mentioning that the positive  bumps in the
 $x<0.1$-region of the NNLO plots of Fig.1  appear
after applying the fine-tuning procedure to
NNLO  coefficients of  $\gamma_{F_3}^{(n)}$ for $n=6,8,10,12$,
which  gives us the  possibility to get reasonable values for
both $\chi^2$ and $\Lambda_{\overline{\rm MS}}^{(4)}$ at NNLO
and beyond.
In otherer words, the $x$-profile of the  twist-4 contribution is related
to the values of the NNLO corrections to $\gamma_{F_3}^{(n)}$
for even $n$. We consider this observation as an additional
argument in favour of getting explicit results for these terms.
\item In general, taking into account systematic experimental uncertainties of the CCFR'97 data for $xF_3$ might make the sinusoidal-type oscillations of
$h(x)$, which demonstrate themselves in the  NNLO and approximate
N$^3$LO fits with smaller number of free parameters $h_i$, less vivid and
more comparable with zero. Indeed, the effective minimization
observed previously in
Refs.~\cite{KKPS2,KPS2}
of the contribution of  twist-4 terms
during less definite than present NNLO fits to $xF_3$ CCFR'97 data
were confirmed in the process of the NNLO DGLAP  fits to
$F_2$ data for charged leptons DIS \cite{Martin:2000gq,Schaefer:2001uh},
which were based on the application of  the
approximate NNLO model for the
DGLAP kernel from Ref.~\cite{vanNeerven:2000ca}.
The observed changes in the $x$-shape of twist-4 contributions $h(x)$
to $xF_3$ and $F_2$ SFs serve as  additional arguments in favour
of high-twist duality, which demonstrates itself through the
interplay between NNLO perturbative QCD corrections and $1/Q^2$ terms.
\item It is interesting to note, that large $\alpha_s$ assosiated
with considerable twist-4 contributions was revealed some time ago
in the process of NLO DGLAP fits of other $\nu N$ DIS data \cite{Zlatev}.
Note, however, that in this work twist-4 contributions were not free,
as in our analysis, but simulated using special model.

\end{enumerate}



\section{Comparison with the results of other NNLO analyses}

In this Section we compare the results
of our studies with the outcomes of other analyses  of
$xF_3$ at the NLO and beyond, peformed independently in the works of
Ref.~\cite{Neerven:2001pe} and Ref.~\cite{Santiago:2001mh}.

In Ref.~\cite{Neerven:2001pe}, to study the evolution of the NS
contributions to  $F_2$ and $xF_3$ up to the approximate N$^3$LO level
of massless QCD,  the NNLO corrections to the DGLAP equation
coefficient functions for $xF_3$ \cite{VZ1} and
the N$^3$LO corrections to definite Mellin moments of NS SFs, obtained
in Refs.~\cite{LNRV,RV}, were combined with the NNLO model
for the NS kernel, previously obtained in
Ref.~\cite{vanNeerven:2000ca}.

\subsection{Comments on estimates of scale-scheme dependence
\\ uncertainties.}

Using the input
\begin{equation}
F_{2,NS}(x,Q_0^2)=xF_3(x,Q_0^2)=x^{0.5}(1-x)^3
\end{equation}
specifying the reference scale $Q_0^2$ through  the
normalization condition
$\alpha_s(\mu_R^2=Q_0^2=30~{\rm GeV}^2)=0.2$  , irrespective
of the order of the expansion and varying the
renormalization scale in the conventional interval
$\frac{1}{4}Q^2\leq \mu_R^2\leq 4Q^2$, the authors of
Ref.~\cite{Neerven:2001pe}
studied the effects  of the scaling violation using  a
definite model  for the $xF_3$ data.
The values of $\alpha_s(30~{\rm GeV}^2)$ for
$f=4$ are given in the second column
of Table 13. However, while analyzing real CCFR'97
$xF_3$ data, we found that scale-dependent uncertainties can be larger.
Indeed, using  those  numbers   from  Table 11,  which are related to
twist-independent fits to the CCFR'97 data,
we obtain the inputs in the third column of Table 13:

\begin{center}
\begin{tabular}{||r|c|c||} \hline
Order
& Values  from Ref.~\cite{Neerven:2001pe} & Our values
 \\ \hline
NLO~~ &   $0.2035^{+0.019}_{-0.011}$    &   $0.2104^{+0.0252}_{-0.0151}$       \\
NNLO~&    $0.1995^{+0.0065}_{-0.0015}$   &  $0.2148^{+0.0097}_{-0.0072}$        \\
N$^3$LO&   $0.2000^{+0.0025}_{-0.0005}$  &  $0.2144^{+0.0058}_{-0.0032}$       \\
\hline
\end{tabular}
\end{center}
{{\bf Table 13.} The comparison of the scale-uncertainties
of $\alpha_s(30~{\rm GeV}^2)$, obtained
in Ref.~\cite{Neerven:2001pe} and in the process of  our studies.}


At the qualitative level both sets of numbers are in agreement
with each other. Moreover, the scale-dependence of both sets
of numbers has the tendency to decrease from NLO up to N$^3$LO.
Definite differences between the central values of $\alpha_s$
may be traced
to the fact that our results for $\alpha_s(30~{\rm GeV}^2)$
correspond to
$\alpha_s(M_Z)\approx 0.118$
irrespective of the order of the expansion (see Eqs.~(39)-(41)), while
the choice $\alpha_s(30~{\rm GeV}^2)=0.2$ in
Ref.~\cite{Neerven:2001pe} corresponds
to a lower value $\alpha_s(M_Z)\approx 0.116$.

On the other hand the scale-dependence of our twist-independent NLO
and NNLO results for $\alpha_s(M_Z)$ (see Eqs.~(39),(40))
is in agreement with the previous estimates
of this kind of theoretical uncertainties, namely
\begin{equation}
\Delta\alpha_s(M_Z)_{NLO}=^{+0.006}_{-0.004}~~~,~~~
\Delta\alpha_s(M_Z)_{NNLO}=^{+0.0025}_{-0.0015}
\end{equation}
obtained in Ref.~\cite{vanNeerven:2000ca} using the model
constructed in this work   for the NNLO NS DGLAP kernel.

As to the application in Ref.~\cite{vanNeerven:2000ca}
of the renormalization-scheme optimization methods of
Refs.~\cite{KS,Kataev:1995rw} for estimating higher-order
corrections (up to N$^4$LO) to the factorization-scheme independent
quantity , defined as
\begin{equation}
\label{K}
{\rm K_n(Q^2)}=-2\frac{{\rm d~ln} M_n^{F_3}}{{\rm d~ln}Q^2}=
\gamma_{F_3}^{(n)}(A_s)-\beta(A_s)\frac{\partial C_{F_3}^{(n)}(A_s)/
\partial A_s}
{C_{F_3}(A_s)}~~~,
\end{equation}
we think that it might give larger theoretical  uncertainties than
those presented in Ref.~\cite{vanNeerven:2000ca}. Indeed, we previously
used this ratio in Ref.~\cite{KPS2} in the process of   the attempt
to perform the massless NNLO fits to CCFR'97 $xF_3$ data
using the effective-charges (ECH)
approach of Ref.~\cite{Grunberg:1984fw} (for the related methods
see Refs.~\cite{Krasnikov:1981rp}-
\cite{Maxwell:2000mm}
and the independent unpublished proposal of Ref.~\cite{Matveev};
for the related phenomenological applications in the  NLO fits
to  the charged leptons DIS SFs data
see Ref.~\cite{Kotikov:1993ht}).
As  was found in Ref.~\cite{KPS2}, the NNLO ECH  fits
to  CCFR'97 $xF_3$ data
face  the problem of the {\it drastical increase} of $\chi^2$
up to the level  of $\chi^2\sim 111/86$. This effect was explained
by the appearance of {\it large} and {\it negative} values of the
NNLO coefficients $\tilde{\beta}_2^{(n)}$ of the ECH $\beta$-functions,
related to the scheme-invariant quantities introduced within the context
of the  principle of minimal sensitivity (PMS)
Ref.~\cite{Stevenson:1981vj}.
Here we are going to demonstrate  explicitly
how this effect  appears,  using the results
presented in Sec.3.

In the case of  $n\geq 2$ the NNLO approximation of the
factorization-scheme independent  kernel
${\rm K_n}$ can be rewritten  in the following form:
\begin{equation}
{\rm R_n}=\frac{K_n}{\gamma_{F_3}^{(0)}(n)}=A_s+d_1(n)A_s^2+d_2(n)A_s^3
\end{equation}
Putting $d_1(n)=d_2(n)=0$ we arrive at the following
renormalization-group equation
\begin{equation}
\left(\mu\frac{\partial}{\partial\mu}+
\beta_{eff}^{(n)}(R_n)\frac{\partial}{\partial R_n}\right)R_n=0~~,
\end{equation}
with the effective $\beta$-function defined as
\begin{equation}
\mu\frac{\partial R_n}{\partial\mu}=\beta_{eff}^{(n)}(R_n)=-2\left(\beta_0 R_n^2+\beta_1 R_n^3+\tilde{\beta}_2^{(n)}
 R_n^4\right)~~,
\end{equation}
where the NNLO coefficient of  Eq.~(47)
is related to the NNLO coefficients $\beta_2$ of the
$\overline{\rm MS}$-scheme $\beta$-function as
\begin{equation}
\tilde{\beta}_2^{(n)} = \beta_2 + \Delta(n)
\end{equation}
with
\begin{equation}
\Delta(n)=\beta_0\bigg(d_2(n)-\Omega_2(n)\bigg)
\end{equation}
and
\begin{equation}
\Omega_2(n)=d_1(n)\left( \frac{\beta_1}{\beta_0}+d_1(n)\right)~~~.
\end{equation}
The similar equations can be derived at the N$^3$LO and beyond
(see Refs.~\cite{KS,Kataev:1995rw}).

The
ECH-inspired estimates proposed in Refs.~\cite{KS,Kataev:1995rw}
work
only in the  case when the differences $\tilde{\beta}_k^{(n)}-\beta_k$
with $k\geq 2$ are small. These conditions  turned out to
be valid for the $e^+e^-$-annihilation Adler
D-function, DIS sum rules in QCD ~\cite{KS} and
$(g-2)_{\mu}$ in QED ~\cite{ Kataev:1995rw} also.
But unfortunately, they are  not working in the case of the
quantity defined by Eq.~(45). Indeed, using the results from
Tables 1,2 we obtain the  numerical expressions
for $\Delta(n)$, which are presented in Table 14:

\begin{center}
\begin{tabular}{||r|c|c|c|c|c|c|c|c|c|c|c|c|} \hline
$n$& 2 & 3 & 4 & 5 & 6 & 7 & 8 & 9 & 10 & 11 & 12 & 13 \\ \hline
$\Delta(n)$& -1976 & -1288 & -1066 & -937 & -851 & -783 &
-730 & -684 & -644 & -606 & -576 & -547 \\ \hline
\end{tabular}
\end{center}
{{\bf Table 14.} The $n$-dependence of
$\Delta(n)=\tilde{\beta}_2^{(n)}-\beta_2$ for $f=4$. }

Comparing now these numbers with the numerical value for $\beta_2$ for $f=4$,
($\beta_2=406.35$), we arrive at the conclusion
that the basic assumption of the ECH-inspired
estimates  of Refs.~\cite{KS,Kataev:1995rw}, namely
$\tilde{\beta}_2^{(n)}\approx \beta_2$,
does  not work at NNLO,
not only in the case  considered previously  of the  correlator of scalar quark currents~\cite{BKM}, but for the factorization-scheme independent
quantity  of Eq.~(46) as well.

Despite the fact that we did not consider in detail
the case of $F_{2,NS}$ SF,
we think
that the estimates presented in Ref.~\cite{Neerven:2001pe}
of  perturbative theoretical uncertainties for the
factorization-scheme independent  kernel  ${\rm K_n}$ for $F_{2,NS}$
made  with the help
of  the ECH and PMS approaches at the NNLO and beyond might be
underestimated  (at least in the  case of $n\leq 13$).
However, the general tendency of the absolute value
of $\Delta(n)$ to decrease with increasing  $n$
might lead to the improvement
of the  situation for a  larger number of moments, which are limited in
Ref.~\cite{Neerven:2001pe} by $n=30$.
This guess can be substantiated  only
after completing  the  {\bf explicit} analytical calculations of
NNLO corrections to DGLAP kernels.

Another interesting subject, related to {\it negative} values
of $\tilde{\beta}_2^{(n)}$,  corresponds to the
appearance of perturbative IRR zeros
of the ECH $\beta$-function at the NNLO, considered  previously as spurious
ones in a number of works
(see Refs.~\cite{Gorishnii:1991zr}-\cite{Gardi:1998rf},~\cite{BKM}).
It should be stressed, that for the quantity  ${\rm K_n}$
these perturbative IRR zeros are  manifesting  themselves
obviously in the case of $n=2$ and less obviously for $n=3$.
In the first case the  critical value of the corresponding effective charge
$(\alpha_s)_{eff}$ is small, namely  $0.4$, while for $n=3$ it is  over
$0.7$, which  is rather
close to the  non-perturbative region. For $n\geq 4$ these
zeros lie  in the typical nonperturbative sector  where $(\alpha_s)_{eff}
\geq 1$. In view of this it might  still be  possible to apply the
ECH or PMS approaches for the analysis of the scheme-dependence
of the NNLO perturbative QCD predictions for $xF_3$ moments with
$n\geq 3$. It could  be of interest to study this problem in the  future.

\subsection{Comments on outputs of the NNLO Bernstein polynomial\\
analyses}

Let us now comment on the comparison of our NLO and NNLO results
for $\Lambda_{\overline{\rm MS}}^{(4)}$ and $\alpha_s(M_Z)$
with the  ones obtained  in another interesting work of
Ref.~\cite{Santiago:2001mh}. The authors
of this work used  the theoretical input identical to ours,
namely the results of the NNLO perturbative
QCD calculations for the
anomalous dimensions and coefficient
functions of odd Mellin moments of $xF_3$, and performed
the NLO and NNLO fits to CCFR'97 data for $xF_3$
with the help of the Bernstein polynomial technique, proposed
in Ref.~\cite{Yndurain:1978wz}. In the process of these
fits, the initial parametrization
 $xF_3(Q_0^2)=Ax^{b}(1-x)^c$, which is similar to Eq.~(6),
was considered.
The initial scales
$Q_0^2=8.75~{\rm GeV}^2$ and $Q_0^2=12~{\rm GeV}^2$
were chosen inside the kinematical region of CCFR'97 data
 $7.9~{\rm GeV}^2\leq Q^2\leq
125.9~{\rm  GeV}^2$, which is only  part of the region used
in our work.

The results of twist-independent fits at the scale $Q_0^2=8.75~{\rm GeV}^2$
obtained in Ref. \cite{Santiago:2001mh} are summarized in the last two columns
of Table 15, where the error bars include statistcial and systematic
experimental uncertainties.

These results can be compared with our outputs
in Table 4, which result from our Jacobi polynomial
twist-4 independent fits of the CCFR'97 data
for $xF_3$ in the kinematical region $Q^2\geq  5~{\rm GeV}^2$ with the cuts
$W > 10 ~{\rm GeV}^2$, $x < 0.7$. It should be stressed that
this region is identical
to the one  studied in the original work of the  CCFR collaboration
~\cite{CCFR2}.
In particular, we are interested in the
following values of
$\Lambda_{\overline{\rm MS}}^{(4)}$
\begin{eqnarray}
\label{our}
LO~~~~~\Lambda_{\overline{\rm MS}}^{(4)}&=&265 \pm 36~{\rm MeV} \\ \nonumber
NLO~~~\Lambda_{\overline{\rm MS}}^{(4)}&=&347 \pm 37~{\rm MeV} \\ \nonumber
NNLO~~~\Lambda_{\overline{\rm MS}}^{(4)}&=&332 \pm 35~{\rm MeV}
\end{eqnarray}
which correspond to the
choice of the initial scale $Q_0^2=8~{\rm GeV}^2$
(note that   the  results of Table 4 demonstrate
that these  values  are
almost independent of  the choice of $Q_0^2$).

It is quite understandable why our results have smaller error-bars:
contrary to the results of Ref. \cite{Santiago:2001mh}
they are defined
by the statistical experimental errors of CCFR'97  data alone.
However, the explanation of the  discrepancies in the
central values of
$\Lambda_{\overline{MS}}^{(4)}$ is not so obvious.

In order to clarify the situation we performed
the Jacobi polynomial fits for two sets of experimental
data from the CCFR'97 collaboration, choosing the same
initial scale {\bf $Q_0^2=8.75~{\rm GeV}^2$} as
in Ref.~\cite{Santiago:2001mh}.

\begin{enumerate}

\item  First, we considered the same data set as in
Ref.~\cite{Santiago:2001mh}, i.e. the kinematical
region   $7.9~{\rm GeV}^2\leq Q^2\leq 125.9~{\rm  GeV}^2$.
\begin{itemize}
\item The comparison of the results of our
twist-4 independent fits with the ones given in Table 3  of
Ref.~\cite{Santiago:2001mh} are presented in Table 15 below.

\begin{center}
\begin{tabular}{||c|c|c|c|c|c|}
\hline
Order & $\Lambda_{\overline{MS}}^{(4)}$ & $\alpha_s(M_Z)$ &
$\chi^2$/nep & $\Lambda_{\overline{MS}}^{(4)}$ \cite{Santiago:2001mh}
 & $\alpha_s(M_Z)$~\cite{Santiago:2001mh} \\
\hline
LO~~& 227$\pm$37 & 0.1309$\pm$0.0037    & 92/74 &  217$\pm$78& 0.130$\pm$0.006
\\
NLO~& 298$\pm$38  & 0.1169$\pm$0.0025 & 76/74 & 281$\pm$57 & 0.116$\pm$0.004
\\
NNLO &303$\pm$38 & 0.1187$\pm$0.0026 & 65/74 & 255$\pm$55 & 0.1153$\pm$0.0041
 \\
\hline
\end{tabular}
\end{center}
{{\bf Table 15.} The values  of
$\Lambda_{\overline{\rm MS}}^{(4)}$ and $\alpha_s(M_Z)$, obtained
with the help of Jacobi and Bernstein polynomial techniques
{\bf $Q_0^2=8.75~{\rm GeV}^2$}
}.
\vspace{0.5cm}

 Notice that the difference between the values of
$\Lambda_{\overline{\rm MS}}^{(4)}$ obtained by the  Jacobi and Bernstein
polynomial techniques are minimized in this case. The LO and NLO
results for $\alpha_s(M_Z)$, as obtained by us, almost coincide with those
taken from Ref.~\cite{Santiago:2001mh}. However, at the NNLO our
value of $\alpha_s(M_Z)$, which is in agreement with the result
of Eq.~(40), is comparable with the similar one given
in Ref.~\cite{Santiago:2001mh} only within the presented experimental
error-bars, which in the latter case also include systematical
experimental uncertainties.
\item To perform a more detailed comparison we
also estimated the uncertainties in the extraction of
$\Lambda_{\overline{\rm MS}}^{(4)}$
at the NNLO, as those considered in Table 4 of
Ref.~\cite{Santiago:2001mh}. The results are given in Table 16.
\begin{center}
\begin{tabular}{||c|c|c|c|c||}
\hline
Sourse~of~errors & $\Lambda_{\overline{MS}}^{(4)}$ & $\Delta
\Lambda_{\overline{MS}}^{(4)}$ &
 $\Lambda_{\overline{MS}}^{(4)}$ \cite{Santiago:2001mh}
 &$\Delta\Lambda_{\overline{MS}}^{(4)}$ \cite{Santiago:2001mh}
  \\
\hline
No~ TMC & 326  &  23    &  298 & 43
\\
HT & 316  & 13 &  270  & 15
\\
$Q_0^2$ to 12 ${\rm GeV}^2$ & 298  & -5  & 263  & -8
 \\
NNLO$^{*}$& 294   & -9  & 209   & -46
 \\
\hline
\end{tabular}
\end{center}

{{\bf Table 16.} Theoretical uncertainties of
$\Lambda_{\overline{\rm MS}}^{(4)}$, obtained
with the help of Jacobi and Bernstein polynomial techniques
{\bf $Q_0^2=8.75~{\rm GeV}^2$.}
}
\vspace{0.5cm}

In Table 16 we mark by the symbol NNLO$^{*}$  the uncertainties
of $\Lambda_{\overline{MS}}^{(4)}$, obtained from the
NNLO fits with $\alpha_s$ defined through its  N$^3$LO expression
(see Eqs.~(14)-(16)). One can see that we obtained twice
as small uncertainties while neglecting TMC and {\rm five} times
smaller effects while introducing the N$^3$LO expression for $\alpha_s$
in the NNLO fits.
It should be stressed that the similar small difference between
the the values of $\Lambda_{\overline{\rm MS}}^{(4)}$  of the
 NNLO and NNLO$^{*}$ fits to the CCFR'97 $xF_3$ data
was already observed in Ref.~\cite{KPS2} for a larger kinematical
region and for different values of $Q_0^2$.

\item At present we are unable to explain the most significant  differences
with the NNLO results of Ref.~\cite{Santiago:2001mh}. We think
that more detailed comparison of
the Jacobi and
Bernstein polynomials approaches at the NNLO
is really on the agenda.
\end{itemize}
\item If, following the CCFR collaboration,
we exclude one data point with $W^2<10~{\rm GeV}^2$
(namely the point with $x=0.65$ and $Q^2=12.6~{\rm GeV}^2$),
which has a large systematical
error, and if we take into account the complete data set with
$Q^2\geq~5~{\rm GeV}^2$ and  the cut $W^2>10~{\rm GeV}^2$,
we reproduce the results of Eq.~(51) with rather small
theoretical uncertainties:
\begin{center}
\begin{tabular}{||c|c|c|c|}
\hline
Order & $\Lambda_{\overline{MS}}^{(4)}$ & $\alpha_s(M_Z)$ &
$\chi^2$/nep
 \\
\hline
LO~~ & 265$\pm$36 & 0.1345$^{+ 0.0031}_{- 0.0033}$    & 113/86
\\
NLO~ & 340$\pm$37  & 0.1193$^{+0.0021}_{0.0023}$  & 87/86
\\
NNLO &333$\pm$36 & 0.1206$^{+0.008}_{-0.0020}$ & 74/86
 \\
NNLO~no TMC &360$\pm$32 &0.1123$^{+0.0018}_{-0.0020}$ & 77/86 \\
NNLO$^{*}$ & 322$\pm$35 &0.1199$^{+0.0021}_{-0.0022}$ & 74/86 \\
\hline
\end{tabular}
\end{center}
{{\bf Table 17.} The results of the fits  to the CCFR'97 data within the
kinematical conditions, used in our work.
{\bf $Q_0^2=8.75~{\rm GeV}^2$.}}
\vspace{0.5cm}

It is worth noting that theoretical uncertainties due to
the omission of TMC and due to the consideration of the N$^3$LO
expression for $\alpha_s$, Eqs.~(14)-(16), in the NNLO fits
remain the same, as in the case of Table 16.

Several additional  comments are now in order
\begin{itemize}
\item In Table 15 and Table 17 the values of $\alpha_s(M_Z)$ were
obtained from Eqs.~(32)-(34) using the matching point $M_5=m_b\approx
4.8$ ${\rm GeV}$. Thus we neglected the uncertainties
due to fixation of $b$-quark threshold ambiguities,
described in Sec. 5.3.
\item Note, however, that our  estimate $(\Delta\alpha_s(M_Z))_{thresh}\approx
\pm 0.0017$ is only slightly  larger than the estimate
$(\Delta\alpha_s(M_Z))_{thresh}\approx 0.0010$, given in
Ref.~\cite{Santiago:2001mh} after application of
a different method for
estimating $b$-quark threshold uncertainties.
\item It should be noted that  contrary to the
analysis in Ref.~\cite{Santiago:2001mh}
we were able to study the scale dependence uncertainties
of our results in the case when the renormalization scale was
taken equal to the factorization scale. The results
of our studies, presented in Sec. 5.3 and Table 11 demonstrate
the following interesting feature: in the case of $k=4$ and HT neglected,
both NLO and NNLO results for $\Lambda_{\overline{\rm MS}}^{(4)}$
are {\bf almost identical} to
the ones obtained in Ref.~\cite{Santiago:2001mh}
in a narrower kinematical region of CCFR'97 $xF_3$ data.
Therefore, a  possible explanation
of the deviations of our results
from the ones in  Ref.~\cite{Santiago:2001mh} might be
related to the fact that scale-dependence ambiguities of the latter
were not studied and might increase the
theoretical uncertainty for $\alpha_s(M_Z)$.

\end{itemize}

\end{enumerate}

\subsection{The model independence of high-twist duality effect}

Despite the fact that in Ref.~\cite{Santiago:2001mh}
the subject related to the inclusion of twist-4 terms
was briefly considered at the NNLO only, it is rather
instructive to perform a similar analysis at the  LO, NLO
and repeat the  NNLO studies using the  Jacobi polynomial approach.
It should be mentioned that in Ref.~\cite{Santiago:2001mh}
the more simple than IRR-model form  of the twist-4  corrections
was used, namely
\begin{equation}
M_{n,xF_3}^{HT}(Q^2)= n\frac{B_2^{'}}{Q^2}M_{n}^{F_3}(Q^2)~~,~~
{\rm with}~~B_2^{'}=a(\Lambda_{\overline{MS}}^{(4)})^2
\end{equation}
Its  coefficient function differs
from  $\tilde{C}(n)$ of Ref.~\cite{DW} (see Eq.~(21)), which
for the moments under consideration has the  following
numerical values: $\tilde{C}(2)=1.6667$, $\tilde{C}(3)=1.6333$,
$\tilde{C}(4)=1.4$, $\tilde{C}(5)=1.0381$, $\tilde{C}(6)=0.5857$,
$\tilde{C}(7)=0.0659$, $\tilde{C}(8)=-0.5063$, $\tilde{C}(9)=-1.1205$,
$\tilde{C}(10)=-1.7689$ and $\tilde{C}(11)=-2.4461$, etc.
Notice that starting from $n$=8, $\tilde{C}(n)$ changes the sign and
in the asymptotic regime  tends to $-n$.
Therefore, it is of interest to investigate  the model-dependence
of the effect observed in Sec.5.1  of high twist duality
using the HT model of Eq.~(53), which is different from the  IRR model
of Ref.~\cite{DW}.

This question was studied by us using the  set of CCFR'97
$xF_3$ data considered in  item (1) of Sec.6.2,
The results, obtained with the help of Jacobi polynomial fits,
are  presented in Table 16,
where $b$-quark threshold uncertainties were not taken into account.
Note that in our fits we considered $B_2^{'}$ as the free parameter,
and then determined the value of the parameter $a$, considered
to be free  in the fits of  Ref.~\cite{Santiago:2001mh}.

\begin{center}
\begin{tabular}{||c|c|c|c|c|c|}
\hline
Order & $\Lambda_{\overline{MS}}^{(4)}$ &
$\chi^2$/nep & $B_2^{'}$ & a &
$\alpha_s(M_Z)$ \\
\hline
LO~~ & 433$\pm$89 & 88/74  & -0.330$\pm$0.126     & -1.76$\pm$0.37 &
0.1471$^{+0.0056}_{0.0063}$
\\
NLO~ & 371$\pm$72   & 75/74 & -0.135$\pm$0.113 & -0.98$\pm$0.57 &
0.1213$^{+0.0039}_{-0.0044}$
\\
NNLO & 316$\pm$51  & 64/74  & -0.031$\pm$0.088 & -0.31$\pm$0.80 &
0.1195$^{+0.0039}_{0.0044}$  \\
\hline
\end{tabular}
\end{center}
{{\bf Table 18.} The results of the  fits to the subset of
 CCCFR'97 data  with
HT contribution, considered  in Ref.~\cite{Santiago:2001mh}.
The initial scale is chosen as $Q_0^2=8.75~{\rm GeV}^2$.}
\vspace{0.5cm}

Comparing now Table 18 with Table 6 of Sec.4 we arrive at the
following observations:

\begin{itemize}
\item In both cases the values of $\Lambda_{\overline{\rm MS}}^{(4)}$ are
almost the same and thus do not depend on the typical structure
of the twist-4 model.
\item At LO and NLO the value of the parameter $B_2^{'}$ is in agreement
with the value of the parameter $A_2^{'}$ of the IRR model and
decreases after NLO effects are taken into account.
\item  At NNLO  the value of the parameter $B_2^{'}$ is a bit larger
than the similar value of $A_2^{'}$, but within error bars both are
compatible with zero.
\item  At NNLO the central value for the parameter $a$ is a bit larger
than its value obtained in  Ref.~\cite{Santiago:2001mh}, but
within existing uncertainties they are compatible.
\item The effect observed in Sec.5.1  of interplay between
perturbative QCD and twist-4 corrections remains also valid in the case
of the choice of twist-4 model given by Eq.~(52).
\end{itemize}

\section{Conclusions}

In this work we used the new perturbative QCD input in the form of
NNLO corrections to the anomalous dimensions and N$^3$LO corrections to
the coefficient functions for odd $xF_3$ moments \cite{RV} to improve
our previous fits to $xF_3$ CCFR'97 data, performed
in the works of Refs.~\cite{KKPS2,KPS2}. We demonstrated that the application
of the smooth interpolation procedure, supplemented
with the fine-tuning of NNLO corrections to $\gamma_{F_3}^{(n)}$ for several
even $n$ gives us the chance to include in the NNLO and approximate
N$^3$LO fits a greater  number of Mellin moments up to $n\leq 13$. The basic
feature which we revealed in the process of our fits is the drastical
reduction
of the scale-dependence uncertainties for $\alpha_s(M_Z)$ at the NNLO
and beyond. The obtained values of $\alpha_s(M_Z)$ turned out to be in
in agreement with the world average value of this parameter, namely
$\alpha_s(M_Z)\approx 0.118$ \cite{Bethke:2000ai,Hinchliffe:2000yq}.
The previously discovered property of interplay between NNLO perturbative
QCD corrections and twist-4 terms is confirmed using different models for the
$1/Q^2$ corrections. The  feature observed in the process
of the NNLO fits with model-independent parametrization of the twist-4
terms is the sinusoidal oscillation of its $x$-shape around zero with
definite positive bumps in the low $x$ region. At present we do not
know whether this typical behaviour can be described by the NNLO
generalization of the IRR model of Ref.~\cite{DW}, or whether  it will
disappear after
taking into account systematic experimental uncertainties.
This   problem can be studied
using the machinery  of the work of Ref.~\cite{Alekhin}, which
 can allow us to fix experimental uncertainties of
$\alpha_s(M_Z)$ and twist-4 terms on more solid ground.

As to the
effective decrease of the twist-4 contributions at the NNLO and beyond,
we are unable to disfavour the possibility that it occurs because
of the use of the CCFR'97 data for $xF_3$, which are still not
precise enough.
Possible future more detailed  DIS $\nu N$ data, which are expected to
be obtained at the
Neutrino factory~\cite{Mangano:2001mj},
 might be useful for clarification  whether it is
reliable  to   detect
more clear
signals from twist-4
contributions at the NNLO level.

On the other hand our NNLO Jacobi polynomial fits revealed the necessity of getting more precise (namely exact) values for the NNLO corrections
to the anomalous dimensions $\gamma_{F_3}^{(n)}$ for even $n$, which
are related to still explicitly uncalculable NNLO corrections to the
kernel
of the DGLAP equation for $xF_3$. Having this information at hand, one might be able to perform NNLO Jacobi polynomial fits avoiding interpolation and fine-tuning procedures, which were used by us
to fix NNLO corrections to  $\gamma_{F_3}^{(n)}$ for even $n$, and
fix still remaining theoretical uncertainties in the $x$-shape
of HT-contribution to $xF_3$, extracted at the NNLO.
More detailed understanding of other  physical  effects  when using
explicit NNLO corrections to the kernels of the DGLAP equations might
be revealed in the process of more detailed comparisons between
different methods, which implement the classical DGLAP solution of differential
equations in the $x$ space and the Jacobi and Bernstein polynomial
techniques.

\section{Acknowledgements}

It is a pleasure to thank
S.I. Alekhin, G. Altarelli, S. Catani, J. Ellis, S. A. Larin, R. Lednicky
A.  Petermann, A.A. Pivovarov,  G. Salam, J. Santiago, D.V. Shirkov,
W.  van Neerven, A. Vogt and F. J. Yndurain
for their interest in our work and constructive questions
and comments.

We are grateful to B.I.  Ermolaev, S. Forte, M. Greco
and A. V. Kotikov  for discussions of
different subjects
related to
the behaviour
of the NS SFs at small $x$.

The essential part of this work was done when one of us (AK)
was visiting Theory Division of CERN.
The work of one of us (GP) was supported by Xunta de Galicia
(PGIDT00PX20615PR) and CICYT (AEN99-0589-C02-02).
From the Russian side the work was supported by RFBI (Grants N 99-01-00091,
00-02-17432). The work of AVS was also supported by INTAS call 2000
(project N 587). While in Russia the work of ALK was done within the framework
of State Support of Leading Scientific Schools (RFBI Grant N 00-15-96626).

\newpage
\begin{center}
Erratum to the revised version of hep-ph/0106221

Improved fits to the $xF_3$ CCFR data at the
next-to-next-to-leading order and beyond.

By   A.~L.~Kataev, G.~Parente and A.~V.~Sidorov,

published in

Physics of Elementary Particles and Atomic Nuclei, vol. 34  (2003)
pp. 43-87

[Physics of  Particles and   Nuclei vol. 34 (2003) pp.  20-46]

\end{center}

The bug crept into the calculations of the numerical values of the
$A_s^3$-coefficients $C_{F_3}^{(3)}(n)$ in the QCD expression for
the  coefficient function  $C_{F_3}^{(n)}(A_s)$ (the  definition
see in Eq.(8))  of the odd Mellin moments of Eq.(1) for $xF_3$
structure function of deep-inelastic neutrino-nucleon scattering
($A_s=\alpha_s/(4/\pi)$) with  $n=1,3,5,7,9,11,13$ and $f=4$
numbers of flavours. This bug resulted from using in the computer
subroutine, which calculated the values for $C_{F_3}^{(3)}(n)$
from the given in  Eq. (22) order $O(A_s^3)$ approximations for
$C_{F_3}^{(n)}$, where instead of $f^2$ in the last terms  $f$ was
typed. These  errors affected also the values of the even
$C_{F_3}^{(3)}(n)$ coefficients, obtained from the explicitly
calculated ones using the smooth interpolation procedure. The
corrected results are given below in the 5-th corrected column of
Table 2 of the paper.
\begin{center}
\begin{tabular}{||r|c|c|c|c|c|c||} \hline
n& $C_{F_3}^{(1)}(n)$ &  $C_{F_3}^{(2)}(n)$ &
$C_{F_3}^{(2)}(n)|_{int}$ & $C_{F_3}^{(3)}(n)|_{int}$ &
$C_{F_3}^{(3)}(n)|_{[1/1]}$ & $C_{F_3}^{(3)}(n)|_{[0/2]}$ \\
\hline
1&  -4     & -52     &  -52       &   -780.6427  &  -676       &  480      \\
2&  -1.778 & -47.472 & (-46.4295) &  (-1206.83008) &  -1267.643  &  174.4079 \\
3&   1.667 & -12.715 &  -12.715   &   -992.198975  &   97.00418  & -47.01328 \\
4&   4.867 &  37.117 & (37.0076)  &  (-269.865143) &   283.0851  &  246.0090 \\
5&   7.748 & 95.4086 &  95.4086   &   851.838501  &   1174.835  &  1013.328 \\
6&  10.351 & 158.2912& (158.4032) &  (2286.68115)  &   2420.569  &  2167.903 \\
7&  12.722 & 223.8978& 223.8978   &   3967.71313   &   3940.284  &  3637.790 \\
8&  14.900 & 290.8840& (290.8421) &  (5844.3042)  &   5678.657  &  5360.371 \\
9&  16.915 & 358.5874&  358.5874  &   7879.04004   &   7601.721  &  7291.305 \\
10& 18.791 & 426.4422& (426.5512) &  (10044.4785)  &   9677.391  &  9391.308 \\
11& 20.544 & 494.1881&  494.1881  &   12319.7676   &   11885.25  & 11633.28  \\
12& 22.201 & 561.5591& (561.2668) &  (14687.1133)  &   14204.22  & 13991.80  \\
13& 23.762 & 628.4539&  628.4539  &   171728.1191   &   16620.99  & 16449.68  \\
\hline
\end{tabular}
\end{center}

{{\bf Table 2.} The  values for  NLO, NNLO, N$^3$LO QCD
contributions to the coefficient functions, used in our fits, and
the results of  N$^3$LO Pad\'e estimates.}

The application of the corrected numbers in the
next-to-next-to-next-to-leading order  Jacobi polynomial fits of
the experimental data of the CCFR-collaboration resulted on slight
decrease of of $N^3LO$ values for $\Lambda_{\overline{\rm
MS}}^{(4)}$ as presented in Tables 6, 11 and 12  by 3 ${\rm MeV}$
only and does not affect any conclusions of the paper.

\begin{thebibliography}{99}
\bibitem{KKPS1}
A. L. Kataev, A. V. Kotikov, G. Parente and A.V. Sidorov,
{\it Phys. Lett.} {\bf B388} (1996) 179.
\bibitem{S1}
A.V. Sidorov, {\em Phys. Lett.} {\bf B389} (1996) 379.
\bibitem{KKPS2}
A.L. Kataev, A. V. Kotikov, G. Parente and A.V. Sidorov,
{\it Phys. Lett.} {\bf B417} (1998) 374.
\bibitem{KPS1}
A. L. Kataev, G. Parente and A. V. Sidorov, {\it
Nucl. Phys.} {\bf A666-667} (2000) 184.
\bibitem{KPS2}
A. L. Kataev, G. Parente and A. V. Sidorov, {\it Nucl. Phys. }
{\bf B573} (2000) 405.
\bibitem{Catani}
S. Catani et al., QCD, CERN-TH/2000-131, In ''Standard Model Physics
(and More) at LHC'' [hep-ph/0005025].
\bibitem{CCFR2}
 CCFR-NuTeV Collaboration, W. G. Seligman et al., {\it Phys. Rev. Lett.}
{\bf 79} (1997) 1213.
\bibitem{PS}
G. Parisi and N. Sourlas, {\it Nucl. Phys.} {\bf B151} (1979) 421.
\bibitem{BLS}
I. S. Barker, C. B. Langensiepen and G. Shaw, {\it Nucl. Phys.}
{\bf B186} (1981) 61.
\bibitem{ChR}
J. Ch\'yla and J. Ramez, {\it Z. Phys.} {\bf C31} (1986) 151.
\bibitem{Kriv1}
V. G. Krivokhizhin, S. P. Kurlovich, V. V. Sanadze, I. A. Savin,
A. V. Sidorov and  N. B. Skachkov, {\it Z. Phys. } {\bf C36} (1987)
51.
\bibitem{Kriv2}
V. G. Krivokhizhin, S. P. Kourlovich, R. Lednicky, S. Nemechek, V. V. Sanadze,
I. A. Savin, V. A. Sidorov and N. B. Schachkov, {\it Z. Phys.}
{\bf C48} (1990) 347.
\bibitem{VZ1}
E. B. Zijlstra and W. L. van Neerven, {\it Phys. Lett.} {\bf B297} (1992) 377.
\bibitem{MV}
S. Moch and J. A. M. Vermaseren, {\it Nucl. Phys.} {\bf B573} (2000) 853;
[hep-ph/9912355].
\bibitem{GL}
V. N. Gribov and L. N. Lipatov, {\it Sov. J. Nucl. Phys.}
{\bf 15} (1972) 438.
\bibitem{L}
L. N. Lipatov, {\it Sov. J. Nucl. Phys.} {\bf 20} (1975) 94.
\bibitem{AP}
G. Altarelli and G. Parisi, {\it Nucl. Phys.} {\bf B126} (1977) 298.
\bibitem{D}
Yu. L. Dokshitzer, {\it Sov. Phys. JETP} {\bf 46} (1977) 641.
\bibitem{LRV}
S.A. Larin, T. van Ritbergen and J. A. M. Vermaseren, {\it Nucl. Phys.}
{\bf B427} (1994) 41.
\bibitem{LNRV}
 S. A. Larin, P. Nogueira, T. van Ritbergen and J. A. M. Vermaseren,
{\it Nucl. Phys.} {\bf B492} (1997) 338;
[hep-ph/9605317].
\bibitem{PKK}
G. Parente, A. V. Kotikov and V. G. Krivokhizhin, {\it Phys. Lett.}
{\bf B333} (1994) 190;
[hep-ph/9405290].
\bibitem{BCDMS}
A.~C.~Benvenuti {\it et al.}  [BCDMS Collaboration],
Phys.\ Lett.\ B {\bf 195} (1987) 91;\\
Phys.\ Lett.\ B {\bf 223} (1989) 485.
\bibitem{RV}
A. R\'etey and J. A. M. Vermaseren,
{\it Nucl. Phys.} {\bf B604} (2001) 281.
[hep-ph/0007294].
\bibitem{Gorishnii:1983su}
S.~G.~Gorishny, S.~A.~Larin and F.~V.~Tkachov,
Phys.\ Lett.\ B {\bf 124} (1983) 217.
\bibitem{Gorishnii:1987gn}
S.~G.~Gorishny and S.~A.~Larin,
Nucl.\ Phys.\ B {\bf 283} (1987) 452.
\bibitem{Chetyrkin:1981qh}
K.~G.~Chetyrkin and F.~V.~Tkachov,
Nucl.\ Phys.\ B {\bf 192} (1981) 159.
\bibitem{LV}
S. A. Larin and J. A. M. Vermaseren, {\it Phys. Lett.} {\bf B259} (1991) 345
\bibitem{Kataev:2000nc}
A.~L.~Kataev, G.~Parente and A.~V.~Sidorov,
to appear in the Proceedings of Quarks-2000 Int. Seminar, Pushkin,
May 2000;
CERN-TH/2000-343,
hep-ph/0012014.
\bibitem{DW}
M. Dasgupta and B.R. Webber, {\it Phys. Lett.} {\bf B382} (1996) 273.
[hep-ph/9604388].
\bibitem{Dokshitzer:1996qm}
Yu.~L.~Dokshitzer, G.~Marchesini and B.~R.~Webber,
Nucl.\ Phys.\ B {\bf 469} (1996) 93
[hep-ph/9512336].
\bibitem{Akhoury:1997rt}
R.~Akhoury and V.~I.~Zakharov, Preprint UM-TH-97-05;
hep-ph/9701378.
\bibitem{Beneke}
M. Beneke, {\it Phys. Rep.} {\bf 317} (1999) 1.
\bibitem{BB}
M. Beneke and V. M. Braun,  Preprint PITHA 00/25, TPR-00-19
[hep-ph/0010208].
\bibitem{Mankiewicz:1997gz}
L.~Mankiewicz, M.~Maul and E.~Stein,
Phys.\ Lett.\ B {\bf 404} (1997) 345
[hep-ph/9703356].
\bibitem{Neerven:2001pe}
W.~L.~van ~Neerven and A.~Vogt,
{\it Nucl. Phys.} {\bf B603} (2001) 42 [hep-ph/0103123].
\bibitem{Yndurain:1978wz}
F.~J.~Yndurain,
Phys.\ Lett.\ B {\bf 74} (1978) 68.
\bibitem{Santiago:2001mh}
J.~Santiago and F.~J.~Yndurain, {\it Nucl. Phys.} {\bf B611} (2001) 447
[hep-ph/0102247].
\bibitem{Alekhin:1999df}
S.~I.~Alekhin and A.~L.~Kataev,
{\it Phys. Lett.}  {\bf B452} (1999) 402
[hep-ph/9812348].
\bibitem{AK2}
S.~I.~Alekhin and A.~L.~Kataev,
{\it Nucl. Phys.}  {\bf A666-667} (2000) 179
[hep-ph/9908349].
\bibitem{Alekhin}
S.~I. Alekhin,
{\it Eur. Phys. J.} {\bf C10} (1999) 395
[hep-ph/9611213].
\bibitem{RVL}
T. van Ritbergen, J.A.M.~Vermaseren and S.A.~Larin,
{\it Phys. Lett.} {\bf B400} (1997) 379;
[hep-ph/9701390].
\bibitem{1}
E.G. Floratos, D.A. Ross and C.T. Sachrajda, {\it Nucl. Phys.}
{\bf B129} (1977) 66; ibid. {\bf B139} (1978) 545 (Err.)
\bibitem{2}
A. Gonz\'alez-Arroyo, C. L\'opez and F.J. Yndurain,
{\it Nucl. Phys.} {\bf B153} (1979) 161.
\bibitem{3}
G. Curci, W. Furmanski and R. Petronzio, {\it Nucl. Phys.} {\bf B175}
(1980) 161.
\bibitem{SY}
J.~Santiago and F.~J.~Yndurain,
{\it Nucl. Phys.} {\bf B563} (1999) 45
[hep-ph/9904344].
\bibitem{SEK}
M.A. Samuel, J.  Ellis  and M. Karliner,
{\it Phys.Rev.Lett.} {\bf 74} (1995)
4380.
\bibitem{KS}
A.~L.~Kataev and V.~V.~Starshenko,
{\it Mod. Phys. Lett.} {\bf A10} (1995) 235
[hep-ph/9502348].
\bibitem{Gardi}
E. Gardi,
{\it Phys. Rev.}  {\bf D56} (1997) 68
[hep-ph/9611453].
\bibitem{PP}
A.~A. Penin  and A.~A.  Pivovarov,
{\it Phys. Lett.}  {\bf B367} (1996) 342
[hep-ph/9506370].
\bibitem{Ellis:1998sb}
J.~Ellis, I.~Jack, D.~R.~Jones, M.~Karliner and M.~A.~Samuel,
{\it Phys. Rev.}  {\bf D 57} (1998) 2665
[hep-ph/9710302].
\bibitem{Mikhailov1}
S.~V.~Mikhailov,
{\it Phys. Lett.} {\bf B431} (1998) 387
[hep-ph/9804263].
\bibitem{Mikhailov2}
S.~V.~Mikhailov,
{\it Phys. Rev.} {\bf D62} (2000) 034002
[hep-ph/9910389].
\bibitem{BKM}
D.~J.~Broadhurst, A.~L.~Kataev and C.~J.~Maxwell,
{\it Nucl. Phys.}  {\bf B592} (2001) 247
[hep-ph/0007152].
\bibitem{Korch}
G.~P.~Korchemsky,
{\it Mod. Phys. Lett.}  {\bf A4} (1989) 1257.
\bibitem{Albino:2000cp}
S.~Albino and R.~D.~Ball, {\it Phys. Lett.} {\bf B513} (2001) 93
[hep-ph/0011133].
\bibitem{Kotikov}
A. V. Kotikov, S. I. Maksimov and V. I. Vovk, {\it Theor. Math. Phys.}
{\bf 84} (1991) 744 [ Teor. Mat. Fiz. {\bf 84} (1991) 101].
\bibitem{Ermolaev:2001sg}
B.~I.~Ermolaev, M.~Greco and S.~I.~Troyan,
{\it Nucl. Phys.}  {\bf B594} (2001) 71
[hep-ph/0009037].
\bibitem{Kirschner:1983di}
R.~Kirschner and L.~N.~Lipatov,
{\it Nucl. Phys.}  {\bf B213} (1983) 122.
\bibitem{PP2}
A.~A.~Penin and A.~A.~Pivovarov,
{\it Phys. Lett.}  {\bf B401} (1997) 294
[hep-ph/9612204].
\bibitem{Sterman:1999yc}
G.~Sterman,
``Recent progress in QCD,'' Plenary talk given at
APS Meeting of the Division of Particle and Fields, Los Angeles, CA,
5-9 January 1999;
hep-ph/9905548.
\bibitem{Dokshitzer:1999ai}
Y.~L.~Dokshitzer,
``Perturbative QCD and power corrections,'' Invited talk at 11th Rencontre
de Blois: Frontiers of Matter, Chateau de Blois, France, 28 June-3 July
1999;\\
hep-ph/9911299.
\bibitem{GB}
D.~J.~Broadhurst and A.~G.~Grozin,
{\it Phys. Rev.} {\bf D52} (1995) 4082
[hep-ph/9410240].
\bibitem{Brodsky:1997vq}
S.~J.~Brodsky, J.~Ellis, E.~Gardi, M.~Karliner and M.~A.~Samuel,
{\it Phys. Rev.}  {\bf D56} (1997) 6980
[hep-ph/9706467].
\bibitem{Brodsky:1983gc}
S.~J.~Brodsky, G.~P.~Lepage and P.~B.~Mackenzie,
{\it Phys. Rev.}  {\bf D28} (1983) 228.
\bibitem{Ball:1995ni}
P.~Ball, M.~Beneke and V.~M.~Braun,
{\it Nucl. Phys.} {\bf B452} (1995) 563
[hep-ph/9502300].
\bibitem{Grunberg:1992ac}
G.~Grunberg and A.~L.~Kataev,
{\it Phys. Lett.}  {\bf B279} (1992) 352.
\bibitem{Rathsman:1996jk}
J.~Rathsman,
{\it Phys. Rev.} {\bf D54} (1996) 3420
[hep-ph/9605401].
\bibitem{vanNeerven:2000ca}
W.~L.~van Neerven and A.~Vogt,
{\it Nucl. Phys.}  {\bf B568} (2000) 263
[hep-ph/9907472].
\bibitem{Chetyrkin:1997sg}
K.~G.~Chetyrkin, B.~A.~Kniehl and M.~Steinhauser,
{\it Phys. Rev. Lett.} {\bf 79} (1997) 2184
[hep-ph/9706430].
\bibitem{Bernreuther:1982sg}
W.~Bernreuther and W.~Wetzel,
{\it Nucl. Phys.}  {\bf B197} (1982) 228
[Erratum-ibid. {\bf B513} (1982) 228].
\bibitem{Larin:1995va}
S.~A.~Larin, T.~van Ritbergen and J.~A.~Vermaseren,
{\it Nucl. Phys.}  {\bf B438} (1995) 278
[hep-ph/9411260].
\bibitem{Blumlein:1999sh}
J.~Blumlein and W.~L.~van Neerven,
{\it Phys. Lett.} {\bf B450} (1999) 417
[hep-ph/9811351].
\bibitem{Shirkov:1997h}
D.~V.~Shirkov, A.~V.~Sidorov and S.~V.~Mikhailov,
hep-ph/9707514.
\bibitem{Shirkov:1992pc}
D.~V.~Shirkov,
{\it Theor. Math. Phys.}  {\bf 93} (1992) 1403
[{\it Teor. Mat. Fiz.}  {\bf 98} (1992) 500].
\bibitem{Shirkov:1994td}
D.~V.~Shirkov and S.~V.~Mikhailov,
{\it Z. Phys.} {\bf C63} (1994) 463
[hep-ph/9401270].
\bibitem{Peterman:1992uj}
A.~Peterman,
CERN-TH.6487/92.
\bibitem{Brodsky:1998mf}
S.~J.~Brodsky, M.~S.~Gill, M.~Melles and J.~Rathsman,
{\it Phys. Rev.}  {\bf D58} (1998) 116006
[hep-ph/9801330].
\bibitem{Jegerlehner:1999zg}
F.~Jegerlehner and O.~V.~Tarasov,
{\it Nucl. Phys.}  {\bf B549} (1999) 481
[hep-ph/9809485].
\bibitem{Kulagin:1998vv}
S.~A.~Kulagin,
{\it Nucl. Phys.} {\bf A640} (1998) 435
[nucl-th/9801039].
\bibitem{Kulagin:1998wc}
S.~A.~Kulagin,
hep-ph/9812532.
\bibitem{Kulagin:2000yw}
S.~A.~Kulagin and A.~V.~Sidorov,
{\it Eur. Phys. J.}{\bf A9} (2000) 261
[hep-ph/0009150].
\bibitem{Webber}
B.~Webber, private communication (1998).
\bibitem{Martin:2000gq}
A.~D.~Martin, R.~G.~Roberts, W.~J.~Stirling and R.~S.~Thorne,
{\it Eur. Phys. J.}  {\bf C18} (2000) 117
[hep-ph/0007099].
\bibitem{Schaefer:2001uh}
S.~Schaefer, A.~Schafer and M.~Stratmann,
{\it Phys. Lett.} {\bf B514} (2001) 284.
[hep-ph/0105174].
\bibitem{Zlatev}
V.~A.~Bednyakov, I.~S.~Zlatev, Y.~P.~Ivanov, P.~S.~Isaev and S.~G.~Kovalenko,
Sov.\ J.\ Nucl.\ Phys.\  {\bf 40} (1984) 494
[Yad.\ Fiz.\  {\bf 40} (1984) 770].
\bibitem{Kataev:1995rw}
A.~L.~Kataev and V.~V.~Starshenko,
{\it Phys. Rev.}  {\bf D52} (1995) 402
[hep-ph/9412305].
\bibitem{Grunberg:1984fw}
G.~Grunberg,
{\it Phys. Rev.} {\bf D29} (1984) 2315.
\bibitem{Krasnikov:1981rp}
N.~V.~Krasnikov,
{\it Nucl. Phys.} {\bf B192} (1981) 497
[{\it Yad. Fiz.}  {\bf 35} (1981) 1594].
\bibitem{Kataev:1982gr}
A.~L.~Kataev, N.~V.~Krasnikov and A.~A.~Pivovarov,
{\it Nucl. Phys.}  {\bf B198} (1982) 508
[Erratum-ibid.\  {\bf B490} (1997) 505]
[hep-ph/9612326].
\bibitem{Dhar:1984py}
A.~Dhar and V.~Gupta,
{\it Phys. Rev.}  {\bf D29} (1984) 2822.
\bibitem{Maxwell:2000mm}
C.~J.~Maxwell and A.~Mirjalili,
{\it Nucl. Phys.}  {\bf B577} (2000) 209
[hep-ph/0002204].
\bibitem{Matveev}
V.~A.~Matveev, private communication (1981).
\bibitem{Kotikov:1993ht}
A.~V.~Kotikov, G.~Parente and J.~Sanchez Guillen,
{\it Z. Phys.}  {\bf C58} (1993) 465.
\bibitem{Stevenson:1981vj}
P.~M.~Stevenson,
{\it Phys. Rev.}  {\bf D23} (1981) 2916.
\bibitem{Gorishnii:1991zr}
S.~G.~Gorishny, A.~L.~Kataev, S.~A.~Larin and L.~R.~Surguladze,
{\it Phys. Rev.} {\bf D43} (1991) 1633.
\bibitem{Chyla:1991ca}
J.~Chyla, A.~L.~Kataev and S.~A.~Larin,
{\it Phys. Lett.}  {\bf B267} (1991) 269.
\bibitem{Vermaseren:1997fq}
J.~A.~Vermaseren, S.~A.~Larin and T.~van Ritbergen,
{\it Phys. Lett.} {\bf B405} (1997) 327
[hep-ph/9703284].
\bibitem{Gardi:1998rf}
E.~Gardi and M.~Karliner,
{\it Nucl. Phys.}  {\bf B529} (1998) 383
[hep-ph/9802218].
\bibitem{Bethke:2000ai}
S.~Bethke,
{\it J. Phys.}  {\bf G26} (2000) R27
[hep-ex/0004021].
\bibitem{Hinchliffe:2000yq}
I.~Hinchliffe and A.~V.~Manohar,
{\it Ann. Rev. Nucl. Part. Sci.}  {\bf 50} (2000) 643
[hep-ph/0004186].
\bibitem{Mangano:2001mj}
M.~L.~Mangano {\it et al.},
``Physics at the front-end of a neutrino factory: a quantitative  appraisal,''
Preprint CERN-TH/2001-131,
[hep-ph/0105155].
\end{thebibliography}
\end{document}